\documentclass{emulateapj}

\slugcomment{Accepted for publication in ApJ}

\newcommand{\sles}{\lower2pt\hbox{$\buildrel {\scriptstyle <}\over
{\scriptstyle\sim}$}} 
\newcommand{\sgreat}{\lower2pt\hbox{$\buildrel {\scriptstyle >}\over
{\scriptstyle\sim}$}} 

\begin{document}

\title{Early optical afterglow lightcurves of neutron-fed Gamma-ray bursts}
\author{Y. Z. Fan$^{1,2,3}$, Bing Zhang$^3$ and  D. M. Wei$^{1,2}$}
\affil{$^1$Purple Mountain Observatory, Chinese Academy of
Sciences, Nanjing 210008, China\\ $^2$National Astronomical
Observatories, Chinese Academy of Sciences, Beijing, 100012, China\\
$^3$Department of Physics, University of Nevada, Las Vegas, NV 89154 }

\begin{abstract}
The neutron component is likely an inevitable ingredient of a
gamma-ray burst (GRB) baryonic fireball, in essentially all progenitor
scenarios. The suggestion that the neutron composition may alter the
early afterglow behavior has been proposed, but there is no detailed
calculation so far. In this paper, within the popular internal shock
scenario of GRBs, we calculate the early optical afterglow lightcurves
of a neutron-fed GRB fireball for different assumed neutron fractions
in the fireball and for both ISM- and wind-interaction models. The
cases for both long and short GRBs are considered. We show that as
long as the neutron fraction is significant (e.g. the number of
neutrons is comparable to that of protons), rich afterglow signatures
would show up. For a constant density (ISM) model, a neutron-rich
early afterglow is characterized by a slowly rising lightcurve
followed by a sharp re-brightening bump caused by collision between
the leading neutron decay trail ejecta and the trailing ion ejecta.
For a massive star stellar-wind model, the neutron-rich early
afterglow shows an extended plateau lasting for about 100 seconds
before the lightcurve starts to decay. The plateau is mainly
attributed to the emission from the unshocked neutron decay
trail. When the overlapping of the initial prompt $\gamma-$rays with
the shocks and the trail is important, as is common for the wind model
and is also possible in the ISM model under some conditions, the IC
cooling effect suppresses the very early optical afterglow
significantly, making  the neutron-fed signature dimmer. For
short GRBs powered by compact star mergers, a neutron-decay-induced
step-like re-brightening is predicted, although the amplitude is not
large.  All these neutron-fed signatures are likely detectable by the
Ultraviolet Optical Telescope (UVOT) on board the {\em Swift}
observatory if GRB fireballs are indeed baryonic and
neutron-rich. Close monitoring of early afterglows 
from 10s to 1000s of seconds, when combined with detailed theoretical
modeling, could be used to potentially diagnose the existence of the
neutron component in GRB fireballs. 
\end{abstract}

\keywords{gamma rays: bursts --- radiation mechanisms: non-thermal ---
shock waves} 

\section{Introduction}
The suggestion that Gamma-ray burst (GRB) fireballs should
contain a good fraction of neutrons has attracted broad attention
recently, since in essentially all progenitor scenarios the neutron
component is likely an inevitable ingredient for a baryonic GRB fireball  
(e.g., Derishev et al. 1999a; Beloborodov 2003a; Pruet et al. 2003).
For instance, core-collapse of massive stars would lead to an outflow
from an iron-rich core with the parameter $\chi$, the neutron-to-proton
number ratio, being $\geq 1$ (e.g., Beloborodov 2003a). In the neutron
star merger model that may be valid for 
short-hard GRBs, one would also expect $\chi\geq 1$. Photo dissociation
during collapse or merger, as well as $n$, $p$ decoupling and
inelastic collisions, would both drive $\chi$ towards unity, although
such an equalization process is likely to remain incomplete (Bachall \&
M\'{e}sz\'{a}ros 2000). Weak interactions induced by
the intense neutrino flux from the central engine can result in 
significant proton-to-neutron conversion, especially if resonant
neutrino flavor transformation takes place (Qian et al. 1993; Fuller
et al. 2000).

Derishev et al. (1999a, 1999b) first investigated the dynamics and the
possible observational signatures of a relativistic neutron-rich
fireball. This was followed by many 
related investigations.
One advantage of the neutron rich
model is that the baryon-loading problem for GRBs can be
ameliorated if significant fraction of baryons confined in the
fireball are converted to neutrons (Fuller et al. 2000). The existence
of the neutron component likely leaves various observational
signatures. For example, the decoupling between the neutron and the
proton components during the early fireball acceleration phase would
give rise to a distinct multi-GeV neutrino emission signature because of
the inelastic neutron-ion collisions in the fireball (Bachall \&
M\'{e}sz\'{a}ros 2000;
M\'{e}sz\'{a}ros \& Rees 2000). Such a GeV neutrino signature is
however not easy to detect in the near future. Lately Fan \& Wei
(2004) suggested that there should be a bright ultraviolet (UV) flash
accompanying a GRB from neutron-rich internal shocks, as
long as the burst is long enough\footnote{As realized in Fan et
al. (2005a), the accelerated electrons are mainly cooled by the
inverse Compton scattering with the initial prompt $\gamma-$ray
photons (Beloborodov 2005), so that the UV flash may be dimmer by 5
mag, but it is still bright enough to be detected by the {\em Swift}
UVOT. The predicted such low energy photon flash accompanying the prompt $\gamma-$ray emission may have already been detected in GRB 041219a (Blake et al. 2005; Vestrand et al. 2005), for which  it is found the optical emission 
during the burst is variable and correlated with the prompt $\gamma-$rays, 
indicating a common origin for the optical light and the $\gamma-$rays (Vestrand et al. 2005). This viewpoint has also been supported by the IR band observation (Blake et al. 2005).}. Such a UV flash is in principle  
detectable by the Ultraviolet-Optical Telescope (UVOT) on board the
{\em Swift} observatory. However, since such signals happen
early (typically $ \geq 14(1+z)$s after the burst trigger), given the
nominal UVOT on-target time (60-100 s), testing this signature
requires an optimized detector configuration. The most promising
approach to test the neutron 
component would be the signatures in the early afterglow phase,
typically tens to hundreds of seconds after the burst trigger, when UVOT is
likely on-target. Such early afterglow neutron signatures have been
suggested earlier (Derishev et al. 1999b; Beloborodov 2003b, hereafter
B03). However, there has been no detailed calculations so far, and
this is the main focus of the current paper.

Our work differs from B03 in three main
aspects: (i) In B03, the neutron shell (N-ejecta) always keeps ahead
of the ion-ejecta (I-ejecta)\footnote{In most cases, ions are
dissociated into protons. Following the convention in the literature,
we call the proton ejecta also as ion ejecta.}, so almost all the
decayed products from 
the N-ejecta are deposited onto the external medium and are used to
accelerate the medium. Such an approximation is only valid
when the burst duration is short enough (e.g. for short bursts).
In this work we show that for typical long GRBs, the separation
between the N-ejecta and the I-ejecta is not clean. The N-ejecta would
partially overlap the I-ejecta until a long distance ${R\sim \rm several
\times R_{\beta}}$, where $R_{\beta}=8\times 10^{15}{\Gamma_{\rm
n,2.5}}{\rm cm}$ is the mean $\beta-$decay ($n\rightarrow
p+e^-+\bar{\nu}_{\rm e}$) 
 radius of the N-ejecta, $\Gamma_{\rm n}$ is the bulk Lorentz
factor (LF) of the N-ejecta, and the convention $Q_{\rm x}=Q/10^{\rm
x}$ is adopted in cgs units here and throughout the text. 
Consequently, a significant fraction of the $\beta-$decay products
are deposited in the I-ejecta rather than in the external medium. So
the neutron-decay trail (i.e., the 
external medium mixed with the $\beta-$decay products) is less
energetic than that suggested in B03. Nonetheless, we confirmed the
conclusion of B03 that the presence of the neutron ejecta
qualitatively changes the afterglow emission properties. (ii) B03
mainly discussed the dynamical evolution of a neutron-fed fireball.
The energy dissipation rates in the neutron front and in the shock
front have been calculated. Although they can delineate the main
characteristics of a neutron-fed fireball, these results are not easy
to be directly compared with the observations. In this work, we extend
B03's investigations to calculate synchrotron radiation both from the
shocks and from the neutron decay trail. We calculated in detail some sample 
early optical lightcurves, which can be directly compared with the future
observations by {\em Swift} UVOT or similar telescopes. (iii) B03
mainly discussed the influence of neutrons on the forward shock (FS)
emission. In this work besides the FS component, we also
explicitly discuss the emission from the reverse shock (RS) region,
which has been widely accepted to play an important role in shaping
early afterglow lightcurves.

The structure of the paper is as follows: In \S{\ref{PhysicsModel}}, we
present the physical picture, including the neutron-rich internal
shocks, the formation of the I-ejecta, the neutron decay trail and
the dynamics of the neutron-fed fireball. In \S{\ref{NumExap}} (for
long GRBs) and \S{\ref{NumExap:Short}} (for short GRBs), we model the
dynamics of neutron-rich systems numerically and calculate the
synchrotron radiation of various emission components. Sample early
optical afterglows for typical parameters are calculated and
presented. For long GRBs (\S{\ref{NumExap}}), the cases for both a
constant-density medium (ISM) and for a stellar-wind medium (wind) are
presented. Our results are summarized in \S{\ref{Dis}} with some
discussions. 

\section{The physical picture}\label{PhysicsModel}
\subsection{n-p decoupling in a neutron-rich fireball} 
In the standard GRB model, a fireball made up of $\gamma$,$e^{\pm}$
and an admixture of baryons is generated by the release of a large
amount of energy $E_{\rm iso}\geq 10^{53} {\rm erg}$ in a region of
$r_{\rm 0}\sim 10^7 {\rm cm}$, where $E_{\rm iso}$ is the total
energy of the burst assuming isotropic energy distribution. Data from
the bursts with known redshifts indicate that a
typical fireball is characterized by a wind luminosity $L\sim
5\times 10^{51} {\rm erg}$ ${\rm s^{-1}}$ and a duration $T_{\rm
90} \sim 50{\rm s}$, measured in the observer frame. Above the fireball
injection radius $r_0$ the bulk Lorentz factor (hereafter LF) $\Gamma$
varies as $\Gamma\sim r/r_{\rm 0}$ initially and saturates when it
reaches an asymptotic value 
$\Gamma_f\leq \eta\sim {\rm constant}$, where $\eta$ is defined as
$\eta=L/\dot{M} c^2$ (e.g. M\'{e}sz\'{a}ros et
al. 1993; Piran et al. 1993).

For an $n$, $p$ fireball, the picture becomes more
complicated\footnote{In this work, we only discuss the purely
hydrodynamic fireball model. If the outflow is magnetohydrodynamic,
the n-p decoupling process is different (Vlahakis et
al. 2003).}. The $n$ and $p$ components are cold in the 
co-moving frame, and remain well coupled until the co-moving nuclear
elastic scattering time $t'_{\rm np}\sim(n'_{\rm p}\sigma_{\rm 0} c)^{-1}$
becomes longer than the co-moving expansion time $t'_{\rm exp}\sim
r/c\Gamma$, where $\sigma_0 \sim 3\times 10^{-26} ~{\rm cm}^2$ is the
pion production cross section above the threshold $\sim
140$ MeV, $n'_{\rm p}$ is the co-moving proton density which according
to mass conservation reads $n'_{\rm p}=L/[(1+\chi)4\pi r^2 m_{\rm
p}c^3\Gamma \eta]$. Therefore the $n$, $p$ decoupling occurs in the
coasting or accelerating regimes depending on whether the
dimensionless entropy $\eta$ is below or above the critical value
$\eta_\pi\simeq 3.9\times 10^2L_{52}
^{1/4}r_{0,7}^{-1/4}[(1+\chi)/2]^{-1/4}$  (Bachall \&
M\'{e}sz\'{a}ros 2000; Beloborodov 2003a).

For $\eta\leq\eta_{\rm \pi}$, both $n$ and $p$ coast with
$\Gamma\simeq \eta={\rm constant}$. For $\eta\geq\eta_{\rm \pi}$, the
condition $t'_{\rm np}\geq t'_{\rm exp}$ is achieved at a radius
$r_{\rm np}/r_{\rm 0}=\eta_{\rm \pi}(\eta_{\rm \pi}/\eta)^{-1/3}$.
While protons are still being accelerated as $\Gamma_{\rm p}\simeq
r/r_{\rm 0}$, the neutrons are no longer accelerated and they coast
at a LF of $\Gamma\simeq\Gamma_{\rm f,n}\simeq2.2 \times
10^2L_{52}^{1/3}r_{\rm 0,7}^{-1/4}[(1+\chi)/2]^{-1/3}\eta_{\rm 3
}^{-1/3}$ (Derishev et al. 1999a). From energy conservation, one gets 
the asymptotic proton Lorentz factor $\Gamma_{\rm
f,p}\simeq\eta(1+\chi)[1-(\chi/[1+\chi])(6/7)(\eta_{\rm
\pi}/\eta)^{4/3}]$  (Bachall \&
M\'{e}sz\'{a}ros 2000).

\subsection{Internal shocks and formation of the I-ejecta}
In the standard fireball model, the long, complex GRBs are powered by
the interaction of proton shells with different LFs. The practical LF
distribution of these shells could be very complicated, and detailed
simulations of these interactions is beyond the scope of this work.
As a simple toy model, here we follow Fan \& Wei (2004) to assume that
LFs of the shells follow an approximate bimodal distribution (Guetta
et al. 2001) with the typical values taken as $\eta_{\rm f}$ (fast
shells) or $\eta_{\rm s}$ (slow shells) with equal probability. 
This simple model is favored for its ability to produce a relatively
high radiation efficiency and a narrow peak energy distribution within
a same burst.
The new ingredient we consider here includes the $n$-component for
both the fast and the slow shells.

Comparing the prompt emission energy and the
afterglow kinetic energy derived from multi-wavelength data fits
(Panaitescu \& Kumar 2002), it is found that a significant fraction of
the initial kinetic energy is converted into internal energy and
radiated as gamma-rays
(Panaitescu $\&$ Kumar 2002; Lloyd-Ronning \& Zhang 2004).
Within the internal shock model, this requires that
the velocity difference between the shells is significant, i.e.
$\eta_{\rm f}\gg \eta_{\rm s}$, and that the two masses are
comparable, i.e. $m_{\rm f}\approx m_{\rm s}$ (Piran 1999). Generally,
$\eta_{\rm s}$ is in the order of tens, and $\eta_{\rm f}$ is in the
order of hundreds. Thus, for slow shells the $n$, $p$ components likely
coast with 
a same LF, i.e. $\Gamma_{\rm s,n}=\Gamma_{\rm s,p}\simeq\eta_{\rm
s}$, while for fast shells the $n$, $p$ components may have different
LFs, which are denoted as $\Gamma_{\rm f,n}$, $\Gamma_{\rm f,p}$,
respectively. 

When an inner faster shell catches up with an outer slower shell,
the ion component merges into a single ion shell. For simplicity
this process is approximated as a relativistic inelastic
collision. Energy and momentum conservations result in
$\Gamma_{\rm m}\simeq [(m_{\rm f,p}\Gamma_{\rm
f,p}+m_{\rm s,p}\Gamma_{\rm s,p}) / (m_{\rm f,p}/\Gamma_{\rm
f,p}+m_{\rm s,p}/\Gamma_{\rm s,p})]^{1/2}$
(Paczy\'{n}ski \& Xu 1994; Piran 1999),
where $\Gamma_m$ is the LF of the merged ion shell.

As shown in Fan \& Wei (2004), the merged ion shells are further
decelerated by the decay products of the slow neutron shells. Given a
small LF of the slow neutron shells, the mean neutron decay radius
is smaller, i.e. $\sim 900 c \Gamma_{s,n} \sim 8\times 10^{14}
\Gamma_{1.5}$ cm. The decay products would be collected by the merged
ion shells at the radii in the range of $R\approx 10^{13}-10^{15}{\rm
cm}$. Shocks are formed, and the ion shells further merge with the
decay products of the slow neutron shell. The  
resulted typical LF of the ion shells would be
\begin{equation}
\Gamma_{\rm I}^0 \approx \sqrt{{(m_{\rm f,p}+m_{\rm s,p})\Gamma_{\rm
m}+m_{\rm s,n}\Gamma_{\rm s,n}\over (m_{\rm f,p}+m_{\rm
s,p})/\Gamma_{\rm m}+m_{\rm s,n}/\Gamma_{\rm s,n}}}~,
\label{Gamma_I}
\end{equation}
which will be regarded as the initial LF of the ion shell (hereafter
called the I-ejecta) whose dynamical evolution we will discuss below. 
The thermal LF of the protons in the shocked region reads
\begin{eqnarray}
\gamma_{\rm th,I}^0 &\approx & {\sqrt{(m_{\rm f,p}+m_{\rm s,p})\Gamma_{\rm
m}+m_{\rm s,n}\Gamma_{\rm s,n}}\over m_{\rm f,p}+m_{\rm s,p}+m_{\rm
s,n}}\nonumber\\
& & \sqrt{(m_{\rm f,p}+m_{\rm s,p})/\Gamma_{\rm
m}+m_{\rm s,n}/\Gamma_{\rm s,n}}-1~,
\label{Gamma_Ith}
\end{eqnarray} 
and the electron synchrotron emission in the region gives rise to a
bright UV flash (Fan \& Wei 2004). 

Notice that the presence of the neutron component does not help to
increase the $\gamma$-ray emission efficiency. Given a same amount of
baryon loading and a same total energy, the LF of the fast 
ion shells is larger than that in the neutron-free model, while the LF
of the slow ion shells ($\sim {\rm tens}$) remains the same. Although
the collision of the 
fast--slow ion shells is more efficient than in the neutron free
model, a significant part of the total energy is carried by 
the fast neutrons shell (i.e. the N-ejecta) most of which is not
translated into the prompt $\gamma-$ray emission. As a result, the
GRB efficiency in the neutron--fed model may be even lower than that 
in the neutron free model. Rossi et al. (2004) gave another argument
on the low radiation efficiency of neutron-rich internal shocks.
It is worth noticing that the internal shock efficiency is also
lowered in a magnetized flow, even magnetic dissipation plays an
essential role (Fan et al. 2004c).

\subsection{The trail of the N-ejecta}

While the fast ion shells are decelerated by the slower ions
(including both the slow ion shells and the decay products of the slow
neutrons), the fast neutron shells (which will be denoted as the
``N-ejecta'') are not. As long as they
move fast enough, say, $\Gamma_{\rm n}\equiv\Gamma_{\rm f,n}>\Gamma_{\rm
I}$ ($\Gamma_{\rm I}$ is the LF of the I-ejecta, considering the
dynamical evolution), the N-ejecta would freely penetrate through all
the ion shells in front of them and would separate from the I-ejecta
more and more. As the result of $\beta$-decay, the mass of the fast
neutrons gradually decreases as 
\begin{equation}
M_{\rm n}(R)=M_{\rm n}^{0}{\rm exp}(-{R/R_\beta}), 
\end{equation}
where $M_{\rm n}^{\rm 0}$ is the original mass of the
total fast neutrons, $R_\beta = \Gamma_n c \tau$ is the
characteristic neutron decay radius, and $\tau \simeq 900$ s is the
rest-frame neutron decay time scale.  If neutrons decay in
the circumburst medium, the $\beta-$decay products $p$ and $e^-$ would
immediately share their momenta with the ambient particles through
two-stream instability (B03). Here we take the number density of the
external medium (in unit of cm$^{-3}$) as
\begin{equation}
n= \left \{ \begin{array}{ll} {\rm const.}, & ({\rm
ISM}),\\ {3.0\times 10^{35}A_*R^{-2}}, & ({\rm wind}),
\end{array}
\label{n0}
   \right.
\end{equation}
where ``ISM'' and ``wind'' represent the uniform interstellar medium
and stellar wind, respectively, $A_*=(\dot{M}/10^{-5}{\rm
M_\odot~yr^{-1}})(v_{\rm w}/10^{3}{\rm km~s^{-1}})^{-1}$, $\dot{M}$ is
the mass loss rate of the progenitor, and $v_{\rm w}$ is the wind
velocity (Dai \& Lu 1998; M\'{e}sz\'{a}ros et al. 1998; Chevalier \&
Li 2000).

B03 discussed a neutron decay ``trail'', which is a mixture of the
ambient particles with the decay products left behind the neutron
front. He then discussed the interaction of the follow-up I-ejecta with
this trail and suggested that the trail would modify the dynamical
evolution of the I-ejecta and hence the afterglow lightcurves. Such a
clean picture is strictly valid for very short bursts, for which the
N-I ejecta separation radius $R_s$ (eq.[\ref{Rs}]) is shorter than
$R_\beta$. For most 
of the long GRBs for which the majority of the afterglow data are
available, not all the decayed mass $|dM_{\rm n}|$ is deposited onto
the circumburst medium. Part of it is deposited onto the I-ejecta
itself. The neutron decay products therefore influence the dynamical
evolutions of both the medium and the ejecta. In this paper, we will
define the ``neutron trail'' only as those decay products that are deposited
onto the circumburst medium. Those deposited onto the I-ejecta
itself will be treated separately. 

\subsection{The dynamical evolution of the system}
With the presence of neutrons, the dynamics of the whole system is far
more complex than the neutron-free case. The physical pictures for
both the ISM and the wind environment are considerably different. 
Below is a qualitative
discussion, and more detailed formulation is presented in 
\S{\ref{NumExap}}. 

For the wind case, the medium is very dense, so the decay products are
immediately decelerated 
significantly, and the resulting bulk LF of the trail is in the order of
tens. The trail is picked up by the I-ejecta quickly. A FS propagates
into the trail and a RS propagates into the I-ejecta. Since
the LF of the FS $\Gamma_{_{\rm FS}}$ is smaller than $\Gamma_{\rm
n}$, the FS never propagates into the wind medium directly,
although at later times the neutron decay rate becomes so low that the
neutron trail essentially does not alter the wind medium. Besides the
emission from both the FS and the RS, for $\chi\sim1$ the unshocked
neutron trail may also 
give non-negligible emission contribution in the optical band. The
thermal LF in the trail is several for typical parameters. Given a
reasonable estimate of the 
local magnetic fields, the typical synchrotron radiation frequency is
around the optical band, so that it could give an interesting
contribution to the optical afterglow emission. This is discussed in
\S\ref{WIND} in detail.

For the ISM case and for $R<{\rm a~few}~R_{\beta}$,
unless $n$ is very large, the inertia of the swept medium is too
small to decelerate the decayed fast neutrons significantly. If the
I-ejecta moves slower than the N-ejecta (which is our nominal case),
the neutron trail most likely moves faster than the I-ejecta,
i.e. $\gamma > \Gamma_{\rm I}$, where $\gamma$ is the LF of the
trail. A gap forms between the trail and the 
I-ejecta. Since the neutron decay rate drops with radius, the trail
LF also decreases with radius. This leads to pile-up of
the neutron trail materials with higher LFs. A
self-consistent treatment of the dynamical evolution of the trail is
complicated. For the purpose of this paper, it is adequate to treat
the fast trail as a global T-ejecta, which in many aspects are
analogous to the I-ejecta. In front of the T-ejecta, there are still
some newly decayed products (we still call these the neutron trail)
behind the N-ejecta front. Even further ahead the N-ejecta is
the unperturbed ISM. The fast moving T-ejecta shocks
into the trail and further into the ISM and gets decelerated. The
Lorentz factor of this forward shock $\Gamma_{_{\rm FS}}$ 
is sometimes larger than the
Lorentz factor of the N-ejecta $\Gamma_{\rm n}$, so that the shock
front is directly in the ISM. 
In any case, the very early afterglow emission is from this
``T-ejecta'' forward shock. The I-ejecta, which lags behind without
significant interaction with the ISM, is not noticeable in the
beginning.  Later, the I-ejecta 
catches up with the decelerated T-ejecta, giving rise to a pair of refreshed
shocks that power strong IR/Optical emission. An energy-injection bump
is expected, which has direct observational consequence. The whole
process is discussed in \S\ref{ISM} in detail.

\section{Early optical afterglow lightcurves: long GRBs}\label{NumExap}

Below we calculate the R-band early afterglow lightcurves, focusing
on the novel properties introduced by the neutron component. In this
section we deal with the traditional long GRBs, those with
$T_{90} > 2$s and believed to be produced during core collapses
of massive stars. Short bursts are discussed in
\S\ref{NumExap:Short}. Two
widely discussed circumburst medium types are ISM and wind, and we
discuss the lightcurves for both cases in turn. It is widely believed
that synchrotron radiation of the shocked electrons 
powers the observed afterglow emission.
The detailed formulae for synchrotron radiation are summarized in
Appendix {\ref{App-A}}. In the text we mainly focus on the
dynamical evolution of the entire system. 

One novel feature of the neutron-fed early afterglow is the emission
from the neutron decay trail, especially in the wind case. The
detailed particle acceleration mechanism as well as the radiation
mechanism in the trail is poorly known. 
In this paper, we adopt two extreme models to treat the trail emission
(see Appendix \ref{App-B} for details). The first treatment is similar
to the shock case, i.e., assuming a power-law distribution of the
electrons. The second treatment is to simply assume a mono-energetic
distribution of the electron energy. We believe that a more realistic
situation lies in between these two cases. In both cases we introduce the
equipartition parameters of electrons and magnetic fields. 

Similar to Fan \& Wei (2004), we take $E_{\rm tot}=2.0\times
10^{53}{\rm ergs}$ (including the total energy of the fast/slow
neutrons and protons at the end of the prompt $\gamma-$ray emission
phase), $\Gamma_{\rm n}=300$, $\Gamma_{\rm s,n}=30$, 
$\Gamma_{\rm m}=200$, $m_{\rm f,p}=m_{\rm s,p}$. In the ISM and the
wind cases, the number density is taken as $n=1$ and
$n=10^{35}R^{-2}$ (i.e. $A_*=1/3$), respectively. The width of the
N-ejecta in the observer frame is taken as $\Delta=10^{12}{\rm cm}$, where
$\Delta=cT_{\rm 90}/(1+z)$ is the width of the ion-ejecta in the rest
frame of the central engine and the observer 
(Correspondingly, $T_{\rm 90}=33(1+z){\rm s}$). In order to find out
how the neutron emission signature depends on the neutron fraction, in
each model we calculate four lightcurves that correspond to the
neutron-to-proton number ratio, $\chi=0.0,~0.1,~0.5,~1.0$, respectively. 

Besides the neutron signature, another important ingredient in the
early afterglow phase is the RS emission. The early dynamical
evolution of a neutron-free shell and its interaction with the
circumburst medium has been
investigated in great detail both analytically (e.g., Sari \& Piran
1995; Kobayashi 2000; Chevalier \& Li 2000; Wu et al. 2003; Z. Li et
al. 2003; Kobayashi \& Zhang 2003b; Fan et al. 2004a; Zhang \&
Kobayashi 2004) and numerically (Kobayashi \& Sari 2000; Fan et
al. 2004b; Nakar \& Piran 2004; Yan \& Wei 2005; Zou et al. 2005).
The early afterglows
of several GRBs (including GRB 990123, GRB 021211 and possibly GRB
021004) have been modeled within the RS emission model (e.g., Sari \&
Piran 1999; M\'{e}sz\'{a}ros \& Rees 1999; Wang et al. 2000; Soderberg
\& Ramirez-Ruiz 2002; Fan et al. 2002; Kabayashi \& Zhang 2003a; Wei
2003; Zhang et al. 2003; Kumar \& Panaitescu 2003; Panaitescu \& Kumar
2004; McMahon et al. 2004). With the presence of neutrons, the
dynamical evolution of the ejecta and the RS emission becomes more
complicated. Below we will formulate the entire process in detail. The
results are reduced to the conventional results when $\chi=0$, but
become more complicated as $\chi$ gets larger.

In this work we assume
that the medium contains electrons and protons only. In reality, at
radii smaller than $10^{16}$cm, the medium may be enriched by
$e^\pm$ pairs created by the interaction of the prompt $\gamma-$rays
with the back-scattered $\gamma$-rays by the 
medium (e.g. Beloborodov 2002 and references therein). The influence
of pairs on the early afterglow lightcurves is neglected in this work
and will be considered elsewhere.

\subsection{Wind case}\label{WIND}

We will first consider the wind model in which no gap is formed
between the I-ejecta and the trail so that
the dynamics is simpler
(see \S{\ref{ISM}} for a detailed treatment for the gap case). The
ejecta-trail interaction can be divided into two phases. (i) Before
the RS crosses the ejecta, $R < R_{\times}$ (where $R_{\rm \times}$ is
the radius at which RS crosses the ejecta), there exist two
shocks, i.e. a FS expanding into the trail, and a RS penetrating
into the I-ejecta. (ii) After the RS crosses the ejecta, $R>R_\times$,
only the FS exists. 

\subsubsection{The dynamical evolution for $R<R_{\times}$}\label{WIND1}

\begin{figure}
\epsscale{1.00}
\plotone{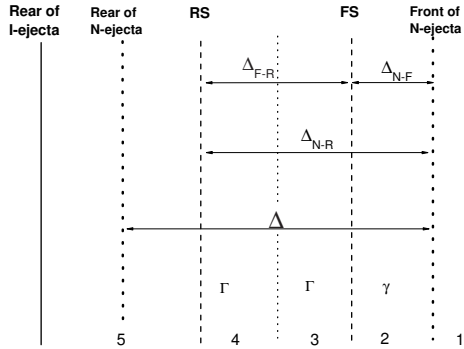}
\caption{A schematic diagram of a neutron-fed long GRB in the wind
interaction case. 
\label{fig:Sch1}}
\end{figure}   

As shown in Figure \ref{fig:Sch1}, there are five regions: (1) the
rest wind medium; (2) the unshocked neutron trail, moving with a LF
$\gamma$; (3) the shocked neutron trail, moving with a LF $\Gamma$;
(4) the shocked I-ejecta, moving with the LF $\Gamma$; and (5) the
unshocked I-ejecta, moving with a LF $\Gamma_{\rm I}$. Notice we have
generally defined the last LF as $\Gamma_{\rm I}$ rather than
$\Gamma_{\rm I}^0$ since the decay products directly deposited onto
the I-ejecta would change $\Gamma_{\rm I}$ from its initial value.

We first treat the dynamics of the neutron trail. The main purpose is
to calculate the velocity of the mixture of the ambient particles
with decay products (i.e. the trail) $\beta(R)=v(R)/c$. Following B03,
we take $dm$, $|dM_{\rm n}|$ as the mass of the ambient medium
overtaken by the N-ejecta and the mass of the decayed fast neutrons,
respectively. For the long bursts we are discussing, only a fraction of
$|dM_{\rm n}|$ is deposited onto the medium (the rest
is deposited onto the I-ejecta). A shock propagates into the trail,
and in the problem what is
relevant is the fraction of the decay products that are deposited onto
the unshocked trail (region 2 in Figure \ref{fig:Sch1}). We denote
this fraction as 
\begin{equation}
f\equiv \min \{1,~{\Delta_{\rm N-F}\over \Delta}\},
\label{f}
\end{equation}
where $\Delta$ is the width of the N-ejecta, which is the same as that of
the I-ejecta, and $\Delta_{\rm N-F}$ is the separation between the
front of the N-ejecta and the front of the FS that propagates
into the neutron decay trail. This is valid for the wind case, in
which $\Delta_{\rm N-F}>0$ is always satisfied. For the ISM case
(\S\ref{ISM}), under certain conditions the FS front leads the
N-ejecta front, and we take $\Delta_{\rm N-F}<0$ following the same
definition here. If this happens, nothing is deposited onto the
unshocked medium, and in our treatment we will define $f=0$
whenever $\Delta_{\rm N-F}<0$ is satisfied.
Notice that by defining $f$, we have assumed that neutrons in the
N-ejecta decay uniformly within the length $\Delta$, which is
justified by the fact $\Delta \ll R$. The same applies to the other
two fraction parameters ($g$, eq.[\ref{g}] and $h$, eq.[\ref{h}] defined
below). 
 
The width $\Delta_{\rm N-F}$ is determined by\footnote{Hereafter we
have adopted $dR \sim c d\hat t$, where $\hat t$ is the time in the
rest frame of the emission source, so that the time interval is
replaced by $dR$.}
\begin{equation}
d\Delta_{\rm N-F} = (\beta_{\Gamma_{\rm n}}-\beta_{\Gamma_{\rm FS}})dR,
\end{equation}
where $\Gamma_{\rm FS}$ is
the LF of the FS
and hereafter we
denote $\beta_{\rm \Gamma_{\rm A}}$ as the dimensionless velocity
corresponding to the LF $\Gamma_{\rm A}$. 

Considering an inelastic collision between the medium (with mass $dm$)
and the decay products (with mass $f |dM_{\rm n}|$), 
energy and momentum conservations give
\begin{eqnarray}
f\Gamma_{\rm n}|dM_{\rm n}|+dm&=&\gamma (\gamma_{\rm
th}+1)(dm+f|dM_{\rm n}|),\nonumber\\
f\Gamma_{\rm n}\beta_{\Gamma_{\rm n}}|dM_{\rm
n}| &=& \gamma \beta_\gamma (\gamma_{\rm th}+1)(dm+f|dM_{\rm n}|)~,
\label{Eq:Bas1}
\end{eqnarray}
where $\gamma_{\rm th}$ is the thermal LF of the mixture
(exclude the rest mass).
With equations (\ref{Eq:Bas1}), we have (see also B03)
\begin{equation}
\beta_\gamma(R)={\beta_{\rm \Gamma_{\rm  n}}\over 1+(\Gamma_{\rm n}
\zeta)^{-1}}, ~~
\gamma(R)={\Gamma_{\rm n}\zeta+1\over (\zeta^2+2\Gamma_{\rm
n}\zeta+1)^{1/2}}; \label{B0301}
\end {equation}
where we have defined a parameter 
\begin{eqnarray}
\zeta(R) &\equiv & f{|dM_{\rm n}(R)|\over dm(R)}=f{M_{\rm n}\over
R_{\rm \beta}}({dm\over dR})^{-1}\nonumber\\
&= & {fM_{\rm n}^0\over
km_{\beta}}({R\over R_{\beta}})^{\rm 1-k}\exp({-{R\over R_{\beta}}}),
\label{B0302}
\end {eqnarray}
which denotes the deposited neutron mass on unit medium mass (notice
that it is slightly different from the same parameter defined in B03
with the correction introduced by the $f$ factor). Here $m_\beta$
is the mass of the medium contained within 
$R<R_\beta$, and $k$ is a parameter denoting the type of the medium,
i.e. $k=1$ for the wind case and $k=3$ for the ISM case. 
For the wind case discussed here, we have 
\begin{equation}
[\zeta(R)]^{\rm wind}={fM_{\rm n}^0\over
m_{\beta}}\exp({-{R\over R_{\beta}}}).
\label{zeta-wind}
\end {equation}

Equations (\ref{Eq:Bas1}) also yields the initial thermal LF of the
neutron trail
\begin{equation}
\gamma_{\rm th}=(\zeta^2+2\Gamma_{\rm
n}\zeta+1)^{1/2}/(1+\zeta)-1.
\label{Gamma_th}
\end{equation}
For $\Gamma_{\rm n}^{-1}\ll \zeta\ll \Gamma_{\rm n}$ (which is valid
for the wind case), the internal
energy of the trail far exceeds its rest mass energy, i.e.,
$\gamma_{\rm th}\gg1$, so that significant radiation is expected. 
When $\zeta>\Gamma_{\rm n}$ (e.g. the ISM case), one has $\gamma_{\rm
th}\ll 1$ so that the trail emission is unimportant.

The total proton number density (in its proper frame) in the unshocked
trail, including the neutron decay products, can be expressed as (B03)
\begin{equation}
n_{\rm tr}=n(1+\zeta)(\zeta^2+2\Gamma_{\rm n}\zeta+1)^{1/2}.
\label{B0303}
\end{equation}

We define $\Delta_{\rm F-R}$ as the width of the shocked regions,
i.e. the distance between the FS and the RS. One then has
\begin{equation}
d\Delta_{\rm F-R}=(\beta_{\rm \Gamma_{\rm FS}}-\beta_{\rm \Gamma_{\rm
RS}})dR, 
\label{Eq:D_FR}
\end{equation} 
where $\beta_{\rm \Gamma_{\rm FS}}$ and $\beta_{\rm \Gamma_{\rm RS}}$
are the dimensionless velocities of the FS and the RS, respectively,
which read (e.g., Sari \& Piran 1995; Fan et al. 2004b)
\begin{eqnarray}
 \beta_{\rm \Gamma_{\rm FS}}&\approx & {\Gamma\beta_{\rm
\Gamma}(4\gamma_{23}+3)-\gamma\beta_{\gamma}\over
\Gamma(4\gamma_{23}+3)-\gamma},\nonumber\\
\beta_{\rm \Gamma_{\rm
RS}} &\approx & {\Gamma\beta_{\rm \Gamma}(4\gamma_{45}+3)-\Gamma_{\rm
I}\beta_{\Gamma_{\rm I}}\over \Gamma(4\gamma_{45}+3)-\Gamma_{\rm I}},
\label{Eq:I1-2}
\end{eqnarray} 
where $\gamma_{23}$ is the LF of the region 2 relative to the region
3, and $\gamma_{45}$ is the LF of the region 5 relative to the
region 4. They are calculated as
\begin{equation}
 \gamma_{23}=\gamma\Gamma(1-\beta_{\gamma}\beta_{\rm
\Gamma}),~~~~\gamma_{45}=\Gamma_{\rm I}\Gamma(1-\beta_{\rm
\Gamma_{\rm I}}\beta_{\rm \Gamma}). 
\label{Eq:I1-3}
\end{equation}

We define another fractional parameter $g$ as the fraction of the
decay products that are deposited onto the shocked region (both
regions 3 and 4 in Figure \ref{fig:Sch1}). We then have
\begin{equation}
g \equiv \left\{
  \begin{array}{@{\,}lc}
    {\Delta_{\rm F-R}}/{\Delta}, & \mbox{for}~ \Delta_{\rm N-R} < \Delta  \\
    1-f, & \mbox{for}~ \Delta_{\rm N-R} > \Delta 
  \end{array}
  \right. ~,
\label{g}
\end{equation}
where $\Delta_{\rm N-R}$ is the separation between the front of the
N-ejecta and the RS front, which is described by 
\begin{equation}
d\Delta_{\rm N-R}=(\beta_{\Gamma_{\rm 
 n}}-\beta_{\rm \Gamma_{\rm RS}})dR. 
\label{Eq:D_NR}
\end{equation}
When $\Delta_{\rm N-R} < \Delta$ the RS has not
passed the end of the N-ejecta, while
when $\Delta_{\rm N-R} > \Delta$, the RS passes through the end of the
N-ejecta, reaching the region where no neutron decay product is being
deposited. 

For $\Delta_{\rm N-R}<\Delta$, some decay products are being deposited
onto the I-ejecta directly (region 5 in Fig.\ref{fig:Sch1}),
which will alter the dynamics of the I-ejecta. We can treat the
process in the way similar to the treatment of the neutron trail onto
the circumburst medium. The difference here is that the I-ejecta
itself is highly relativistic. The energy and momentum conservations
are more analogous to those to describe internal shock collisions. 
Unlike the circumburst medium case in which the trail LF only depends
on $R$, one needs two parameters to describe the I-ejecta dynamics.
Besides the radius $R$, one needs to specify the location of the
I-ejecta element in the I-ejecta proper. We parameterize this by the
distance from the initial I-ejecta front, which is denoted as
$\Delta_1$ so that $0<\Delta_1 <\Delta$ is satisfied. 
After some algebra, the LF of the I-ejecta layer with the coordinate
$\Delta_1$ at a particular $R$ could be written as
\begin{equation}
\Gamma_{\rm I}(\Delta_1,R)=\sqrt{{\bar\zeta \Gamma_{\rm
n}+(1+\gamma_{\rm 
th,I}^0)\Gamma_{\rm I}^0 \over \bar\zeta /\Gamma_{\rm
n}+(1+\gamma_{\rm 
th,I}^0)/\Gamma_{\rm I}^0}}, 
\label{Eq:I1-5}
\end{equation}  
and the corresponding thermal LF reads
\begin{equation}
\gamma_{\rm th,I}(\Delta_1,R)=\sqrt{{(\bar\zeta \Gamma_{\rm n}+\Gamma_{\rm
I}^0)[\bar\zeta /\Gamma_{\rm n}+(1+\gamma_{\rm th,I}^0)/\Gamma_{\rm
I}^0]\over (1+\bar\zeta )^2}}-1,
\label{Eq:I1-6}
\end{equation}
where 
\begin{equation}
\bar\zeta = \bar\zeta(\Delta_1,R) \equiv \frac{M_{\rm
n}^0}{M_{\rm I}^0} \left[1-\exp\left(-\frac{R_1}{R_\beta}\right)
\right],
\label{zetabar}
\end{equation}
and $M_{\rm I}^0$ is the initial mass of the I-ejecta.  
The parameter $\bar{\zeta}$ is analogous to the $\zeta$ parameter
(eq.[\ref{B0302}]) for the trail treatment, which denotes the average
decayed neutron mass on unit I-ejecta mass. The characteristic radius
$R_1$ depends on whether or not the I-ejecta layer (with coordinate
$\Delta_1$) overlaps the N-ejecta as the reverse shock sweeps the
layer (so that the layer no longer belongs to the region 5). If
the layer still overlaps with the N-ejecta as it is swept by the RS,
through out the period when the layer is in the region 5 there has
been always neutron decay products deposited onto the layer. In such a
case, one simply has $R_1=R$. For those layers that the end of the
N-ejecta has passed over earlier so that there is no neutron decay
deposition at the RS sweeping time, one has to record the decayed
neutron amount at the final moment with N-ejecta overlap. In this
case, one has $R_1\approx 2(\Gamma_{\rm I}^0\Gamma_{\rm
n})^2(\Delta-\Delta_1)/(\Gamma_{\rm 
n}^2-{\Gamma_{\rm I}^0}^2)$. 

The RS crossing radius $R_\times$ can be calculated according to
\begin{equation}
\int^{R_{\times}}_0 d\Delta_{\rm I-R}=\Delta,
\end{equation}
where $\Delta_{\rm I-R}$ is the distance between the I-ejecta front and
the RS assuming no shock compression. It is determined by 
\begin{equation}
d\Delta_{\rm I-R}=(\beta_{\rm \Gamma_{\rm I}}-\beta_{\rm \Gamma_{\rm
RS}})dR.
\label{Eq:I1-8}
\end{equation} 

With the above preparation, one can delineate the dynamical evolution
of the system for $R<R_\times$ using the following equations.
After some algebra, the energy conservation of the system yields
\begin{eqnarray}
(M+U/c^2)d\Gamma=(\gamma-\Gamma \gamma_{23})(1+\gamma_{
\rm th})dm_{\rm tr}+
(\Gamma_{\rm I} \nonumber\\
-\Gamma\gamma_{45})(1+\bar\zeta
)(1+\gamma_{\rm th,I})dm_{\rm I}+g(\Gamma_{\rm n}-\Gamma)|dM_{\rm n}|,
\label{Eq:I1-9}
\end{eqnarray}
in deriving which, the small thermal energy generated by the
deposition of the neutron decay products into the shocked region
has been neglected. Here $U$ is the thermal energy generated in the
two shock fronts, which can be calculated by 
\begin{eqnarray}
dU/c^2&= & (1-\epsilon_1)[\gamma_{23}(1+\gamma_{\rm th})-1]dm_{\rm
tr}\nonumber\\
&+&(1-\epsilon_2)(1+\bar{\zeta})
[\gamma_{45}(1+\gamma_{\rm
th,I})-1]dm_{\rm I}, 
\label{Eq:I1-10}
\end{eqnarray}
where $\epsilon_1$ and $\epsilon_2$ are the radiation efficiency (see
the Appendix \ref{App-A} for definition). The differential increase of
the trail mass, the I-ejecta mass (just initially) and the total mass
in the shock region can be expressed as
\begin{eqnarray}
dm_{\rm tr} & = & [1+\zeta(R)]4\pi n R^2dR, \label{mtr}\\
dm_{\rm I} & = & M_{\rm I}^0d\Delta_{\rm I-R}/\Delta, \\
dM & = & dm_{\rm tr}+(1+\bar\zeta )dm_{\rm I}+g|dM_{\rm n}|,
\label{Eq:I1-11}
\end{eqnarray} 
Combining equations (\ref{Eq:D_FR}-\ref{Eq:I1-11}) with the well known
radius--time relation 
\begin{equation}
dR=\beta_{\rm \Gamma_{\rm FS}}c~dt/(1-\beta_{\rm \Gamma_{\rm FS}}),
\label{Eq:I1-12}
\end{equation} 
one can calculate the dynamical evolution for $R<R_{\times}$ (see our
numerical example in \S\ref{WindL}). Hereafter $t=t_{\rm obs}/(1+z)$
denotes the observer's time 
corrected for the cosmological time dilation.

\subsubsection{The dynamical evolution for $R>R_{\times}$}
After the RS crosses the I-ejecta, only the FS exists. Equations
(\ref{Eq:I1-9}, \ref{Eq:I1-10}, \ref{Eq:I1-11}) can be simplified as
\begin{equation}
(M+U/c^2)d\Gamma=(\gamma-\Gamma \gamma_{23})(1+\gamma_{
\rm th})dm_{\rm tr}+g(\Gamma_{\rm n}-\Gamma)d|M_{\rm n}|,
\label{Eq:I2-1}
\end{equation}
\begin{equation}
dU/c^2=(1-\epsilon_1)[\gamma_{23}(1+\gamma_{\rm th})-1]dm_{\rm tr},
\label{Eq:I2-2}
\end{equation}
\begin{equation}
dM=dm_{\rm tr}+g|dM_{\rm n}|.
\label{Eq:I2-3}
\end{equation} 

\begin{figure}
\epsscale{1.00}
\plotone{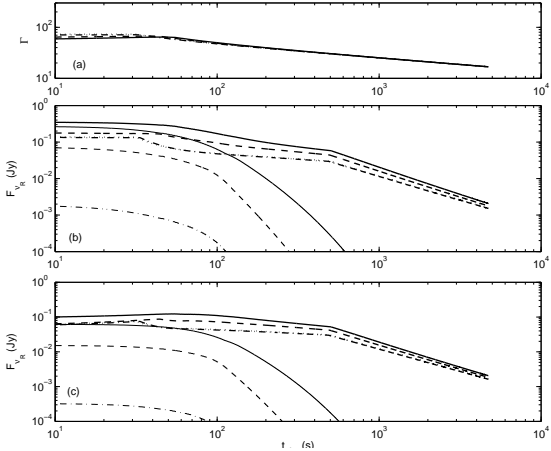}
\caption{
The early optical afterglow lightcurves of a neutron-fed long GRB in
the wind interaction case. (a) The dynamical evolution of the LF of
the shocked region as a function of time. (b) R-band
lightcurves, with the IC cooling effect due to the prompt
$\gamma-$rays interacting with the shocked regions being
ignored. Thick lines  include contributions from all emission
components, including the FS, 
RS and the neutron decay trail. Thin lines are for trail emission
only (for a power-law energy distribution of the electrons, see
Appendix \ref{App-B} for details).  The
dotted, dash-dotted, dashed and solid lines represent
$\chi=0.0,~0.1,~0.5,~1.0$ respectively. Following input parameters are
adopted in the calculations: $E_{\rm tot}=2.0\times 10^{53}{\rm
ergs}$, $\Delta=10^{12}$cm, $z=1$ [i.e. $D_{\rm L}=2.2\times
10^{28}{\rm cm}$, which corresponds to the standard
$(\Omega_m,\Omega_\Lambda)=(0.3,0.7)$ $\Lambda$CDM cosmological
model], $\Gamma_{\rm n}=300$, $\Gamma_{\rm m}=200$, 
$\Gamma_{\rm s,n}=30$, and $n=10^{35}{\rm cm^{-3}}R^{-2}$
(i.e. $A_*=1/3$), respectively. The parameters $\epsilon_{\rm e}=0.1$,
$\epsilon_{\rm B}=0.01$ and $p=2.3$ are adopted for 
the FS and RS shocks as well as the trail. (c) R-band
lightcurves, but the IC cooling effect due to the prompt $\gamma-$rays
overlapping with the shocked region and the trail has been taken into
account. The averaged $\gamma-$ray luminosity is taken as
$L_\gamma=10^{51}{\rm ergs}~{\rm s}^{-1}$. Other parameters and line
styles are the same as those in (b).  
\label{fig:WIND}}
\end{figure} 

\subsubsection{The impact of the initial prompt  
$\gamma-$rays on electron cooling}

As pointed out by Beloborodov (2005), if the RS emission
overlaps with the initial prompt $\gamma-$ray emission, the
 cooling of the shocked electrons is likely
dominated by the inverse Compton (IC) scattering off the initial
prompt $\gamma-$rays. This is a very common case for the
wind-interaction case, as has been studied in detail in Fan et
al. (2005a), who discuss the interesting sub-GeV photon emission and
high energy neutrino emission due to the overlapping effect. In the
ISM case, usually the prompt $\gamma-$ray emission has crossed the RS
region before the RS emission becomes important, so that the above
overlapping effect is not as important as in the wind
model\footnote{We can understand this as follows.
For illustration, we take the neutron-free model ($\chi=0$). The
ejecta are decelerated significantly at the radius $R_{\rm dec}\sim
10^{17}{\rm cm}~E_{\rm iso,53.3}^{1/3}n_0^{-1/3}\Gamma_{\rm
m,2.3}^{-2/3}$, at which the rear of the prompt $\gamma-$ray emission
leads the rear of the ejecta at an interval  
$\sim R_{\rm dec}/2\Gamma_{\rm m}^2\sim 1.2\times 10^{12}{\rm
cm}E_{\rm iso,53.3}^{1/3}n_0^{-1/3}\Gamma_{\rm m,2.3}^{-8/3}$, 
which is usually larger than (or at least comparable to) the width of
the fireball $\Delta\sim 10^{12}{\rm cm}$. As a result, for the typical
parameters taken in this work, the IC cooling due to the overlapping
in the ISM case is not as important as in the wind model.}. We have
performed calculations of the overlapping effect in the ISM model as
well, and found that the resulted early optical afterglow lightcurves  
are nearly unchanged for typical parameters. So we do not
present the result in Figure \ref{fig:ISM}. Nonetheless, we would
like to point out that for some GRBs born in the ISM environment whose
duration is long enough and/or whose bulk LF is large enough, the
overlapping effect should be also very important. This has been the
case of GRB 990123 (Beloborodov 2005), which has a long ($T_{90}\sim
120{\rm s}$) duration and a large LF ($\sim 1000$). 

Similar to Fan et al. (2005a), we assume that the internal shock
efficiency is $r\simeq 0.2$ and define the IC cooling  
parameter $Y_{\rm s}\equiv U_{\rm \gamma}/U_{\rm 
B}$. Here $U_{\gamma}\simeq L_{\gamma}/4\pi R^2\Gamma^2c$ is the
MeV photon energy density in the rest frame of the shocked region,
and $U_{\rm B}=B^2/8\pi\simeq (\Gamma_{\rm
I}^0/\Gamma)^2 \epsilon_B n_4 m_p c^2$ is the 
co-moving magnetic energy density in the same region, where
$L_\gamma\simeq r c E_{\rm tot}/(1-r)\Delta$ is the luminosity of the 
$\gamma-$ray burst, and $n_4 \approx
E_{\rm tot}/4\pi R^2 {\Gamma_{\rm I}^0}^2 m_{\rm p} c^2\Delta $ is the
co-moving density of the un-shocked ejecta.
After some simple algebra, we derive $Y_{\rm s}\approx r /[(1-r
)\epsilon_{\rm B}]$. Notice that this is only valid when the
Klein-Nishina correction is unimportant, i.e. $x\ll 1$, where $x\equiv
\epsilon_{\gamma}\gamma_{\rm e}/\Gamma m_{\rm e}c^2 \sim \gamma_{\rm
e}/\Gamma$, $\epsilon_\gamma \sim  m_{\rm e}c^2$ is the energy of the
prompt $\gamma-$rays, $\gamma_{\rm e}$ is the random LF of the
emitting electrons. More generally, we should make the Klein-Nishina
cross section correction, i.e. $\sigma_{\rm IC}=A(x)\sigma_{\rm T}$, 
where $A(x)\equiv {3\over 4}[{1+x\over x^3}\{{2x(1+x)\over 1+2x}-{\rm
ln}(1+2x)\}+{1\over 2x}{\rm ln}(1+2x)-{1+3x\over (1+2x)^2}]$, with the
asymptotic limits $A(x)\approx 1-2x+{26x^2\over 5}$ for $x \ll 1$, and
$A(x)\approx {3\over 8}x^{-1}({\rm ln}2x+{1\over 2})$ for $x \gg 1$
(e.g. Rybicki \& Lightman 1979). Notice that such a correction is
necessary since the target photons, i.e. the initial prompt
$\gamma-$rays are usually energetic enough in the rest frame of the
electrons. As a result, for the electrons accelerated by the reverse
shock, for $x \gg 1$, the Compton parameter reads (see Footnote 1 of Fan et al. [2005a]  for 
derivation)
\begin{equation}
Y_s \approx {A(x)r \over x (1-r) \epsilon_B}
\label{Y_s}
\end{equation}
Usually, the magnetic field densities in the FS and the RS are nearly
the same (e.g. Sari \& Piran 1995), so Eq. (\ref{Y_s}) also applies to
the FS shock region. The overlapping of the initial prompt $\gamma-$rays
with the FS emission lasts until $\int{(1-\beta_{\rm FS})dR}=\Delta$. 

Similarly, for the electrons accelerated in the neutron trail, we can
also introduced the IC cooling parameter $Y_T\approx A(x_1)U_{\gamma,T}/(x_1U_{B'_{\rm tr}})$ for $x_1\gg 1$, where $U_{\rm \gamma,T}\approx L_{\gamma}/4\pi
R^2\gamma^2c$, $B'_{\rm tr}$ is calculated by Eq. (\ref{B_tr}), and
$x_1\equiv \epsilon_\gamma \gamma_{\rm e}/\gamma m_{\rm e}c^2\sim
\gamma_{\rm e}/\gamma$. 

When the prompt $\gamma-$rays overlap the shocked regions and the trail,
to calculate the cooling frequency, the $Y$ parameter in
Eqs. (\ref{gamma_ec}), (\ref{gamma_emt}) and
(\ref{gamma_ect}) should be replaced by $Y_{\rm s}$ (for electrons
accelerated from the shock) or $Y_{\rm T}$ (for electrons accelerated
in the neutron trail).

\subsubsection{Model lightcurves}\label{WindL}

In Figure \ref{fig:WIND} we present the dynamical evolution of the
system (Figure \ref{fig:WIND}(a)) and the early optical afterglow
lightcurves in Figs.\ref{fig:WIND}(b) and (c). In
Fig.\ref{fig:WIND}(b) the IC cooling effect due to
the initial prompt $\gamma-$ray overlapping with the shocked region and
the trail has been ignored. In Fig.\ref{fig:WIND}(c), this IC cooling
contribution is explicitly taken into account. The 
neutron-to-proton number density ratio $\chi$ is adopted as 
$\chi=0.0, 0.1, 0.5$ and 1, respectively.

As shown in Fig.\ref{fig:WIND}(a), for $R<R_{\times}$, $\Gamma$ is
nearly a constant. This is consistent with the familiar result in the
neutron-free wind model (e.g., Chevalier \& Li 2000; Wu et al. 2003;
Kobayashi \& Zhang 2003b; Fan et al. 2004b; Yan \& Wei 2005; 
Zou et al. 2005). The main reason is that
the density of the N-ejecta also drops as $\propto R^{-2}$, and that
the fraction of the decayed neutrons essentially remains the same
during the shock crossing process. The latter effect is manifested by
the fact that $R_\times$ is comparable to $R_\beta$ for the current
nominal parameters, so that the neutron decay rate [$\propto
\exp(-R/R_\beta)$] remains essentially the same for $R<R_\times$.

The dynamical evolution of the shocks presented
here [e.g. Fig.\ref{fig:WIND}(a)] look different from the one
presented in Fig.1 of B03, which shows that the dynamics of a
neutron-fed fireball is very different from the neutron-free one. 
This is because B03 presented the dynamical evolution of the trail in
the radiative regime, 
i.e., the total thermal energy of the trail is radiated promptly
before the trail is picked up by the I-ejecta. Here we assume an
electron equipartition parameter $\epsilon_{\rm e}=0.1$ in the neutron
trail\footnote{As noted in B03, the energy dissipation mechanism in
the neutron decay trail could be fundamentally different from that
in the collisionless shock. Nonetheless, for known mechanisms of
particle acceleration in a proton-electron plasma, the fraction of
thermal energy given to electrons is usually significantly
smaller than that given to protons. We therefore suspect that
$\epsilon_{\rm e}\sim 0.1$ may be reasonable.}. For such a
small $\epsilon_{\rm e}$, even in the fast cooling regime, only 0.1 of
the total thermal energy is radiated, and the system could be still
approximated as an adiabatic one. Our results indicate that as long as
the trail is in the quasi-adiabatic regime, the bulk of the energy is
still dissipated in shocks, and the existence of the neutrons does not
influence the dynamical evolution of the system significantly (see
also B03 for similar discussions).

In any case, we include the emission contribution from the neutron
trail by assuming a power-law distribution of the electrons, and find
that it gives an interesting signature when $\chi$ is 
close to unity. In Fig.\ref{fig:WIND}(b) we present the lightcurves for
different $\chi$ values, with the IC cooling by the initial prompt
$\gamma-$rays ignored. The thick lines are the synthesized
lightcurves including the contributions from the FS, RS and the
neutron trail. In order to characterize the function of the trail, we
plot the trail emission separately as thin lines. We can see that the
early afterglow for $\chi=1$ shows a bright 
plateau lasting for $\sim 100$s, and the main contribution is from the
trail. For $\chi \leq 0.5$, an early afterglow plateau with a shorter
duration is still evident, but it is mainly due to the contribution of
the RS emission\footnote{The lightcurve temporal index before shock
crossing is $\sim 0$ rather than 1/2 discussed in other papers
(e.g. Chevalier \& Li 2000; Wu et
al. 2003; Kobayashi \& Zhang 2003b; Fan et al. 2004b; Yan \& Wei 2005;
 Zou et al. 2005). The main reason
is that for the 
nominal parameters in this paper, the optical band is above both
$\nu_m$ and $\nu_c$ for the RS emission, $F_{\nu_{\rm R}}\propto
t^{\rm -(p-2)/2}$, which is very flat.}. Comparing the lightcurves 
with small $\chi$'s (e.g. 0.0 and 0.1) to those with large $\chi$'s
(e.g. 0.5 and 1.0), one can see that the early afterglow intensity is
much stronger for high-$\chi$ case, although the dynamical evolution
of the I-ejecta is rather similar [Fig.\ref{fig:WIND}(a)]. The
reason is that the trail deposits an electron number density about 10
times of that in the wind medium, so that the total number of the
emitting electrons in the shocked region is greatly increased.

In Fig.\ref{fig:WIND}(c) we present the lightcurves by taking into
account the IC cooling effect due to the initial prompt $\gamma-$rays
overlapping with the shocked regions and the trail. Line styles are the
same as in Fig. \ref{fig:WIND}(b). It is apparent that the very early
R-band emission, including the FS emission, the RS emission and the
trail emission, has been suppressed significantly. This means that the
IC cooling effect is very essential in shaping the early afterglow
lightcurves in the wind case. After the rear of initial prompt
$\gamma-$ray front has crossed the FS front, the FS emission becomes
rather similar to that of the Fig.\ref{fig:WIND}(b). 

\begin{figure}
\epsscale{1.0}
\plotone{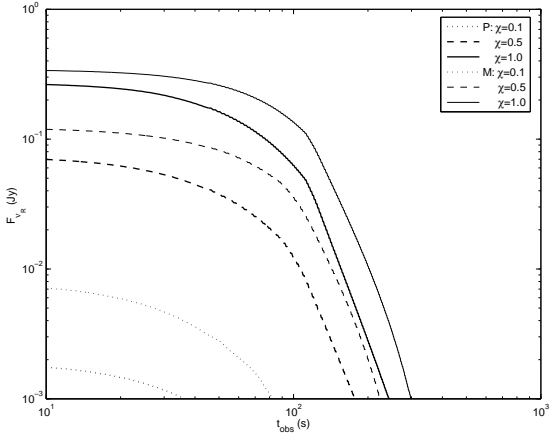}
\caption{
The trail emission lightcurve in the wind interaction case, with the
IC cooling effect due to the initial prompt $\gamma-$rays overlapping
with the trail ignored.  The
dotted, dashed and solid lines represent $\chi=0.1,~0.5,~1.0$
respectively. The thick lines (marked by P) represent the emission by
assuming a power law distribution of electrons, and thin lines (marked
by M) represent the emission by assuming a mono-energetic
distribution. The initial input parameters are the same as those
presented in Figure \ref{fig:WIND}(b) caption.
\label{fig:Trail}}
\end{figure}

In the wind case, the difference between the neutron-rich lightcurve 
and a neutron-free lightcurve is only marginal (both Fig.\ref{fig:WIND}(b)
and Fig.\ref{fig:WIND}(c)). One potential tool to 
diagnose the existence of the neutron component is to search for the
trail emission component. This is in principle not straightforward
since there are many uncertainties in categorizing the trail emission
(Appendix \ref{App-B}). In Figure \ref{fig:Trail} (with the IC
cooling effect ignored), we calculate the 
expected trail emission for two extreme models for electron energy
distribution, i.e. a power law model and a mono-energetic model. A more
realistic model should lie in between these two models. Figure
\ref{fig:Trail} indicates a fortunate fact that the differences
between the two extreme models are far from large, although the
mono-energetic model predicts a stronger R-band trail emission. This
gives us confidence that the crude treatment of the trail emission
in Fig.\ref{fig:WIND} gives a first-order presentation of the
reality. Another remark is that different electron energy
distributions result in different trail emission spectrum. In
particular, a mono-energetic or thermal-like distribution gives rise to
a much lower emissivity at high frequencies. Even in the optical
band, multi-color observations at early times can be used to diagnose
the existence of the neutron trail. For example, if the trail electron
energy distribution is thermal-like or in a form largely deviated from
the simple power-law, multi-color observations around 100s with the
UVOT on board {\em Swift} (there are 6 colors in the 170-650 nm band
for UVOT) can reveal interesting clues of the existence of neutrons in
the fireball. 

\subsection{ISM case}\label{ISM}

We now turn to discuss the ISM-interaction for long GRBs. The main
characteristic of the ISM case is that the trail typically moves
faster than the I-ejecta, so that a gap forms between the I-ejecta and
the T-ejecta (the trail ejecta). The latter interacts with the
ISM and dominates the early afterglow. Later the I-ejecta catches up
and produces an energy injection bump signature. Below we will
calculate the typical lightcurves with the nominal parameters adopted
in this paper. Two stages will be considered separately, i.e. (1) when
the T-ejecta is separated from the I-ejecta and dominates the
afterglow emission, and (2) when the I-ejecta collide onto the
decelerated T-ejecta. 
Except otherwise stated, the notations adopted in this section are the
same in \S\ref{WIND} when the wind case is discussed.

\subsubsection{The dynamical evolution of T-ejecta before I-ejecta
collision}

The $\zeta(R)$ parameter (eq.[\ref{B0302}]) in the ISM case reads
\begin{equation}
[\zeta(R)]^{\rm ISM} ={fM_{\rm n}^0\over
3 m_{\beta}}({R\over R_{\beta}})^{-2}\exp({-{R\over R_{\beta}}}),
\end {equation}
For $\chi=1$, the nominal parameters are $\Gamma_{\rm
I}^0=100$, $\Gamma_{\rm n}=300$, $\gamma_{\rm th,I}^0=0.433$ and
$\zeta(R)=2.85 \times 10^4 
f (R/R_\beta)^{-2}\exp(-R/R_\beta)$.  The N-ejecta leads the
I-ejecta, and completely separates from the
I-ejecta at 
\begin{equation}
R_{\rm s}\simeq \frac{\Delta}{\beta_{\Gamma_{\rm n}}-\beta_{\Gamma_{\rm I}^0}}
\simeq \frac{2{\Gamma_{\rm I}^0}^2\Gamma_{\rm 
n}^2\Delta}{(\Gamma_{\rm n}^2-{\Gamma_{\rm I}^0}^2)}.
\label{Rs}
\end{equation} 
As long as $f\exp(-R/R_\beta)(R/R_\beta)^{-2}>2.34\times
10^{-3}$, one has $\gamma(R)>\Gamma_{\rm I}^0$ (eq.[\ref{B0301}]). The
I-ejecta lags behind the trail, and a gap forms between the I-ejecta
and the trail. This happens for $\chi>{\rm several}~\times 0.1$, when
the decay products deposited onto the ISM have a large enough momentum
to drag the trail fast enough to lead the I-ejecta. With equation
(\ref{Rs}), we can  
estimate how many decay products have been deposited into the I-ejecta directly
\begin{eqnarray}
M_{\rm n,d} &\simeq & M_{\rm n}^0\int_0^{\rm R_{\rm s}}(1-{R\over R_{\rm
s}})\exp(-{R\over R_{\rm \beta}}){dR\over R_\beta} \nonumber\\
&=& [1+{\exp({\rm -a})\over a}-{1\over a}]M_{\rm n}^0,
\end{eqnarray} 
where $a=R_{\rm s}/R_\beta$. For $a>2$, we have $M_{\rm n,d}>0.5M_{\rm
n}^0$, i.e., most of decay products have been  
deposited into the I-ejecta.

For the trail, the layers from the behind moves faster than the
leading layers, since $[\zeta(R)]^{\rm ISM}$ decreases with $R$. This leads to
pile-up of the trail materials. The faster trail materials would shock
into the slower trail materials in front. In the following treatment,
we approximately divide the trail materials into two parts. The fast
trail (including forward shocked trail) is denoted as the T-ejecta,
while the upstream unshocked part is still called the trail. The
location of the shock is roughly defined by requiring the bulk LF of
the T-ejecta (which is calculated through energy and momentum
conservation) is larger than $2\gamma$, so that the relative LF
between the two parts exceeds 1.25. Since the trail is hot with a
relativistic temperature, the upstream local sound speed $c/\sqrt{3}$
corresponds to a local sound LF $\simeq\sqrt{3/2}\sim 1.22$. So our shock
forming condition is self-consistent.
The shock front moves faster than the T-ejecta, and $\beta_{\rm
\Gamma_{\rm FS}}$ may exceed $\beta_{\rm \Gamma_{\rm n}}$. If this
persists long enough, the N-ejecta lags behind the FS front. The
decay products deposit into the shocked material directly rather than
mixing with the ISM. The shock propagates into the ISM
directly. In such a case, the dynamics is simplified.  

\begin{figure}
\epsscale{0.8}
\plotone{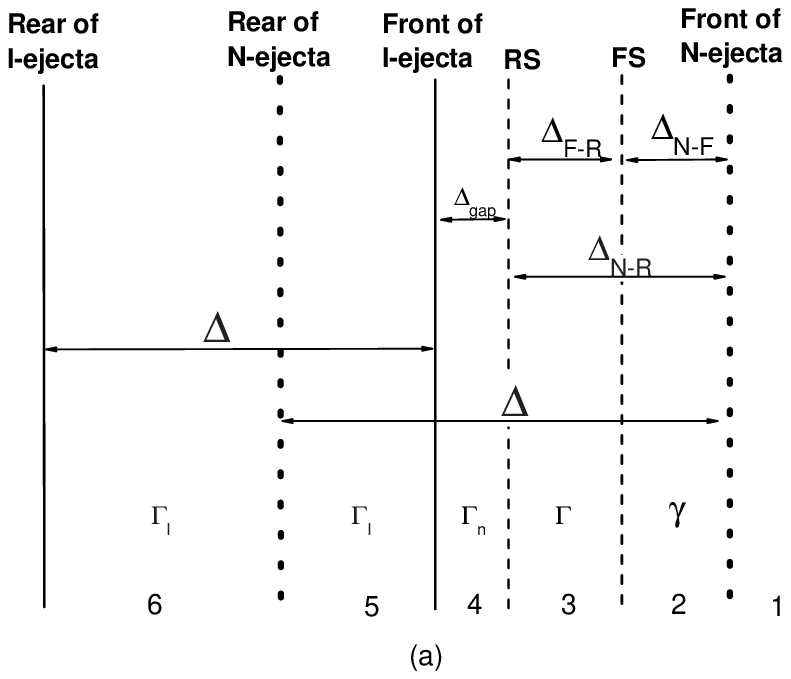}
\plotone{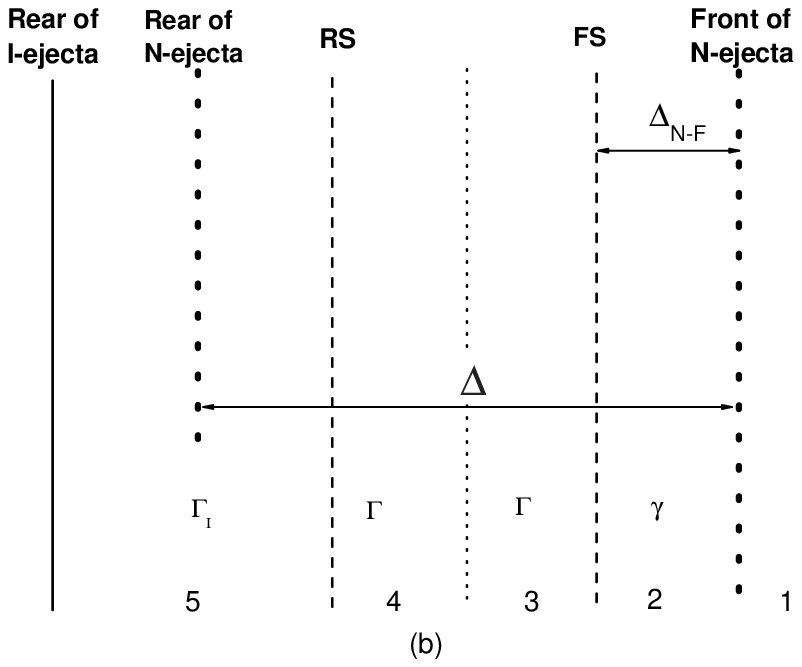}
\caption{The schematic diagram of a neutron-fed long GRB in the
ISM-interaction case. (a) When the T-ejecta (region 3) is separated from
the I-ejecta and interacts with the trail or ISM.  (b) When the I-ejecta
collides into the T-ejecta producing refreshed shocks.
\label{fig:Sch2}}
\end{figure}  

Before collision, there are six regions in the whole system
(see Fig.\ref{fig:Sch2}(a) for illustration): (1) the rest
ISM; (2) the unshocked trail moving with $\gamma$; (3) the T-ejecta
(including the shocked trail between the FS and the RS) 
moving with $\Gamma$; (4) the gap between the T-ejecta rear and
the I-ejecta front, in which the decayed protons move with $\sim
\Gamma_n$; (5) the overlapped region of the I-ejecta and
the N-ejecta; and (6) free moving I-ejecta. 

The width of the T-ejecta (i.e., region 3), $\Delta_{\rm F-R}$ is
still governed by equation (\ref{Eq:D_FR}), but now
$\beta_{\Gamma_{\rm RS}}=[\Gamma
\beta_{\Gamma}(4\gamma_{34}+3)-\Gamma_{\rm n}\beta_{\Gamma_{\rm n}}]/ 
[\Gamma (4\gamma_{34}+3)-\Gamma_{\rm n}]$, where
$\gamma_{34}=\Gamma\Gamma_{\rm n}(1-\beta_{\Gamma}\beta_{\Gamma_{\rm
n}})$ is the LF of region 4 relative to region 3.  
The fraction of the decay products deposited into the T-ejecta
directly, $g$, is still defined by equations (\ref{g}) and
(\ref{Eq:D_NR}), in which we take $f=\min\{1,\Delta_{\rm 
N-F}/\Delta\}$, and $f=0$ when $\Delta_{\rm N-F}$ is negative.

Energy conservation of the T-ejecta interacting with the regions 2 and
4 results in\footnote{The thermal energy generated in the decay products
depositing into T-ejecta is small and is therefore ignored here.} 
\begin{eqnarray}
(U/c^2+M)d\Gamma =  (\gamma-\Gamma\gamma_{23})(1+\gamma_{\rm
th})dm_{\rm tr}\nonumber\\
+g(\Gamma_{\rm n}-\Gamma)|dM_{\rm n}|
+(\Gamma_{\rm
n}-\Gamma \gamma_{34})dm_{\rm gap}, 
\label{Eq:II1-3}
\end{eqnarray}
where $U$ is the thermal energy generated in both the FS and RS
fronts, which is 
described by 
\begin{eqnarray}
dU/c^2&=&(1-\epsilon_1)[\gamma_{23}(1+\gamma_{\rm th})-1]dm_{\rm
tr}\nonumber\\
&+&(1-\epsilon_2)(\gamma_{34}-1)dm_{\rm gap}, 
\end{eqnarray}
$dm_{\rm gap}$ is the differential mass swept by the RS from the gap,
which can be approximated by 
\begin{equation}
dm_{\rm gap}=l M_{\rm gap}{d\Delta_{\rm N-R}\over \Delta_{\rm gap}},
\label{Eq:m_gap}
\end{equation}
where $l=1$ for $\Delta_{\rm N-R}\leq \Delta$ and $l=0$ otherwise (see
equation (\ref{Eq:D_NR}) for the definition of $\Delta_{\rm
N-R}$). This step function is used to characterize whether there is
decay product falling into the gap region. $M_{\rm gap}$ is the total
mass contained in the region (4), which satisfies  
\begin{equation}
dM_{\rm gap}={l\Delta_{\rm gap}\over \Delta}|dM_{\rm n}|-dm_{\rm gap}.
\label{Eq:M_gap}
\end{equation}
 The width of the gap $\Delta_{\rm gap}$ (i.e., the 
separation of the T-ejecta rear and the I-ejecta front),
is governed by  
\begin{equation}
d\Delta_{\rm gap}=\left\{
  \begin{array}{@{\,}lc}
    (\beta_{\Gamma_{\rm RS}}-\beta_{\Gamma_{\rm I}})dR, & \mbox{for}~
\Delta_{\rm N-R} < \Delta,  \\ 
    (\beta_{\Gamma}-\beta_{\Gamma_{\rm I}})dR, & \mbox{for}~
\Delta_{\rm N-R} > \Delta. 
  \end{array}
  \right. 
\label{Eq:D_gap}
\end{equation}  

 The total mass in the T-ejecta
can be calculated through 
\begin{equation}
dM=dm_{\rm tr}+g|dM_{\rm n}|+dm_{\rm gap},
\label{Eq:II1-4}
\end{equation} 
where $dm_{\rm tr}$ is still defined by equation (\ref{mtr}). Notice
that the expression of $dm_{\rm tr}$ is reduced to $dm_{_{\rm ISM}}$ when
$f=0$ is satisfied, i.e. there is no decay products deposited onto the
unshocked medium. This usually does not happen in the wind case, but
happens sometimes in the ISM case.

During this stage, the I-ejecta evolves independently. Because of the
deposition of decay products, the I-ejecta is somewhat accelerated. 
To estimate this effect, we assume that the I-ejecta moves with a
uniform LF $\Gamma_{\rm I}$. In principle, the part that
overlaps with the N-ejecta (so that deposition of the decay products
is possible) should move slightly faster than the rest part, and
within the decay-area different layers may move with slightly
different LFs. But such an effect is unimportant for our further
discussions about the interaction between the I-ejecta and the
T-ejecta. 
Energy and momentum conservations then yield 
\begin{equation}
(M_1+U_1/c^2)d\Gamma_{\rm I}=[(\Gamma_{\rm n}-\Gamma_{\rm
I})-\Gamma_{\rm I}(\gamma_{\rm rel}-1)]h|dM_{\rm n}|,
\label{Eq:II1-5}
\end{equation} 
where
\begin{equation}
h=\max \{0,1-{\Delta_{\rm gap}+\Delta_{\rm F-R}+\Delta_{\rm N-F}\over
\Delta}\}, 
\label{h}
\end{equation}
and $\gamma_{\rm rel}=\Gamma_{\rm I}\Gamma_{\rm n}
(1-\beta_{\Gamma_{\rm I}}\beta_{\Gamma_{\rm n}})$ is the relative LF
between the N-ejecta and the I-ejecta, $U_1$ and $M_1$ are determined
by $dU_1/c^2=(\gamma_{\rm rel}-1)h|dM_{\rm n}|$ and  $dM_1=h|dM_{\rm
n}|$ (Originally, $M_1=M_{\rm I}^0$), respectively. We can then solve
$\Gamma_{\rm I}$ at any time, 
which is adopted as the input parameter in the next section.
 
\subsubsection{The collision between the I-ejecta and the T-ejecta}   

At the beginning of the evolution, one has $\Gamma>\Gamma_{\rm I}$,
and the I-ejecta lags behind further and further. However, the
T-ejecta is decelerated by the trail and the ISM continuously, so that
$\Gamma$ decreases with radius. When $\Gamma=\Gamma_{\rm I}$ the
separation between the I-ejecta and the T-ejecta is the largest. Later
one has $\Gamma<\Gamma_{\rm I}$, and the gap between the two ejecta
shrinks, until eventually the I-ejecta catches up with the T-ejecta. 
This is a standard energy injection process, and a pair of
refreshed shocks are formed, i.e. a refreshed FS (RFS) expanding into
the hot T-ejecta, and a refresh RS (RRS) penetrating into the
I-ejecta. The detailed treatment of such a process has been presented
before (e.g. Kumar \& Piran 2000; Zhang \& M\'{e}sz\'{a}ros
2002). In particular, Zhang \& M\'{e}sz\'{a}ros (2002) considered the
interactions between three media, i.e. the ISM, the initial fireball
shell and an injected shell. They considered three emitting shocks,
including the leading forward shock and the pair of the refreshed
shocks. Their analysis is pertinent to treat the current problem.
As shown in Zhang \& M\'{e}sz\'{a}ros (2002), the
emission powered by the RFS is unimportant in the optical
band, (see the short dashed line in their Fig. 4b). Since we are
mainly focused on the optical lightcurves in this paper, that result
leads to great simplification of our treatment. In our calculations,
only the RRS and the FS are taken into account. The simplified system includes 5
parts (see Figure \ref{fig:Sch2}(b) for illustration): (1) the unshocked
ISM, (2) the unshocked trail (this region may merge to region 1), (3)
T-ejecta and shocked trail, (4) shocked I-ejecta, and (5) 
unshocked I-ejecta. Notice that we no longer separate the unshocked
I-ejecta into two regions 
(whether overlap with the N-ejecta), since we treat the unshocked
I-ejecta as a whole with a same LF $\Gamma_{\rm I}$.

Energy conservation of the the system (T-ejecta, the shocked
trail and the shocked I-ejecta) yields
\begin{eqnarray}
(U_2/c^2+M_2)d\Gamma=(\gamma-\Gamma\gamma_{23})(1+\gamma_{\rm
th})dm_{\rm tr}\nonumber\\
 +(\Gamma_{\rm I}-\Gamma \gamma_{\rm 45})
(1+\gamma_{\rm
th,I})dm_{\rm I}+g(\Gamma_{\rm n}-\Gamma)|dM_{\rm n}|, 
\label{Eq:II2-1}
\end{eqnarray}
where $\gamma_{23}$ and $\gamma_{\rm 45}$ are the relative LFs between
region 2 and 3, and region 4 and 5, respectively (eq.[\ref{Eq:I1-3}]),
$U_2$ is the thermal energy generated in the shocks, 
which is determined by
\begin{eqnarray}
 dU_2/c^2&=&(1-\epsilon_1)[(1+\gamma_{\rm
th})\gamma_{23}-1]dm_{\rm tr}\nonumber\\
&+&(1-\epsilon_2)[(1+\gamma_{\rm
th,I})\gamma_{\rm 45}-1]dm_{\rm I},
\end{eqnarray}
and $M_2$ is the total mass in the shocked region
\begin{equation}
dM_2=dm_{\rm tr}+dm_{\rm I}+g|dM_{\rm n}|.
\end{equation}
Here $\gamma_{\rm th,I}=U_1/M_1c^2$, 
$dm_{\rm I}\approx M_1d\Delta_{\rm I-R}/\Delta$, $d\Delta_{\rm
I-R}=(\beta_{\Gamma_{\rm I}}-\beta_{\rm \Gamma_{\rm RS}})dR$ 
(eq.[\ref{Eq:I1-8}]), and $M_1$ is the total I-ejecta mass upon
collision (according to eq.[\ref{Eq:II1-5}]). The 
velocity of the RRS reads $\beta_{\rm \Gamma_{\rm RS}}\approx
[\Gamma\beta_{\rm \Gamma}(4\gamma_{\rm 45}+3)-\Gamma_{\rm
I}\beta_{\rm \Gamma_{\rm I}}]/[\Gamma(4\gamma_{\rm 45}+3)-\Gamma_{\rm I}]$.

After the reverse shock crosses the I-ejecta, the I-ejecta and
the T-ejecta merge into one single shell, and the influence of the
neutron component essentially disappears. The later dynamical evolution of
this merged shell is the same as that in the neutron-free model, which
has been studied in many publications (e.g. Huang et al. 2000;
Panaitescu \& Kumar 2002). We do not discuss it further in this paper.

\subsubsection{Model Lightcurves}\label{ISML}

\begin{figure}
\epsscale{1.00}
\plotone{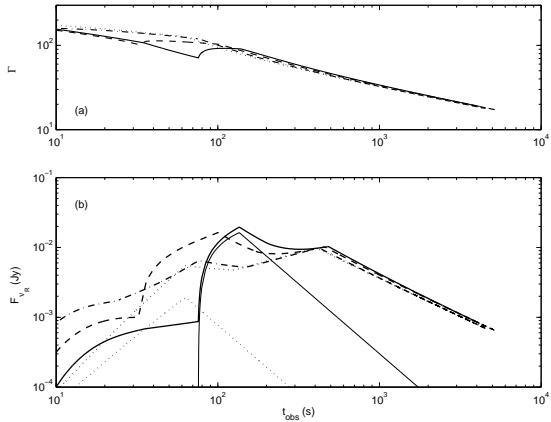}
\caption{
The early optical afterglow lightcurves of a neutron-fed long GRB in
the ISM interaction case. (a) The dynamical evolution
of the region shocked by FS as a function of time. The dotted,
dash-dotted, dashed, and solid lines represent
$\chi=0.0,~0.1,~0.5,~1.0$ respectively. (b) R-band lightcurves. Line
styles are the same as in (a). Thick lines represent the total early
R-band lightcurves, while the thin lines are for RS emission only. 
For clarity, only the RS emission for $\chi=0$ and the RRS emission for $\chi=1$
cases are plotted.
The initial parameters are the same as those listed in the caption of
Figure \ref{fig:WIND}, except that $n=1{\rm cm^{-3}}$ is adopted.
\label{fig:ISM}}
\end{figure}

The dynamical evolution of the system (LF of the T-ejecta) and the
R-band lightcurve in the ISM-interaction case are calculated for
$\chi=$0, 0.1, 0.5, 1.0, respectively (see Figure
\ref{fig:ISM})\footnote{Our dynamical evolution of the system for the
ISM case is also 
different from that of B03. The discrepancy is mainly due to different
assumptions involved. In this work, we assume that the I-ejecta  moves
slower than the N-ejecta, while in B03, they are assumed to be the 
same.}. In 
the cases of $\chi=0.5$ and $\chi=1$, there is a gap between the
I-ejecta and the T-ejecta, and a bump is evident in both the
$\Gamma-t_{\rm obs}$ and $F_{\nu_{\rm R}}-t_{\rm obs}$ diagrams. This
bump is the signature for the I-ejecta and T-ejecta collision. When
$\chi \leq 0.1$, this signature is not significant, mainly because the
gap is not well-developed.
The trail is picked up by the I-ejecta continuously, so that the
treatment is similar to the one to calculate the wind case (see
\S{\ref{WIND}}). The R-band lightcurve for $\chi=0.1$ is
rather similar to the $\chi=0$ case, except that it is relatively
brighter at earlier times (e.g. before the RS crosses the I-ejecta). 
The reason is that even in the $\chi=0.1$ case, the trail is much
denser than the ISM, so that there are more electrons in the forward
shock front that contribute to synchrotron radiation. 
The early R-band flux for $\chi=0.5$ and $\chi=1$ is also different from
the case of $\chi=0$. This is because the early afterglow for a
high-$\chi$ case is powered by the T-ejecta, which has a smaller mass in
the beginning than the I-ejecta. The T-ejecta soon enters the
deceleration phase, so that $\Gamma$ starts to drop with time at a
very early epoch (see Figure \ref{fig:ISM}(a)). In our calculation, it
is found out that for high-$\chi$ case, the initial RS that crosses
the T-ejecta is too weak for any observational consequence. 
For a comparison, for the low-$\chi$ case ($\chi \leq 0.1$), the more
massive I-ejecta is decelerated slowly, 
and it is not significantly decelerated within 100 seconds (see Figure
\ref{fig:ISM}(a)). In the meantime, a strong RS crosses the I-ejecta,
whose emission is included in our calculations. This keeps a higher
level of the early afterglow emission. In all the cases, the rising
behavior of the early afterglow is because the typical synchrotron
frequency is above the band. For the high-$\chi$ case, the obvious
bump signature is the joint contribution from the FS and the RRS, and
latter is the dominant component in the optical band. For
illustrative purpose, in Figure \ref{fig:ISM}(b), besides the total
lightcurve, we also plotted the RRS contribution in the $\chi=1.0$
case, and the initial RS contribution in the $\chi=0.0$ case.

An interesting question is to explore how to use the lightcurve to
diagnose the existence of the neutron component for the ISM case.
For example, the initial rising lightcurve of a neutron-free fireball
is steep, while for a neutron-fed fireball it is much milder because
of the contribution of the shocked trail emission.
The sharp bump in the early afterglow stage is a prominent
signature. However, it suggests that an energy injection event
happens, not necessarily a proof of the existence of the neutron
component. Nonetheless, the neutron-fed model gives a natural
mechanism for post energy injection, and it suggests that an early
bump is common for long GRBs with ISM-interaction as long as $\chi$ is
reasonably large, say $>0.5$. If future Swift observations reveal that
an early bump is common, it may suggest that the GRB fireballs are
baryonic with a large fraction of neutrons. It is worth mentioning
that for the neutron-free model, an early afterglow bump is also
expected due to the transition from the RS emission to the FS emission
(e.g. Kobayashi \& Zhang 2003a; Zhang et al. 2003). Nonetheless, the
bump due to energy injection (presumably powered by the collision
between the I-ejecta and the T-ejecta) could be well-differentiated. 
For example, the injection bump is achromatic, while the FS peak is chromatic,
caused by crossing of the typical frequency into the band.
Finally, the RS emission 
for $\chi=0$ is weaker than the RRS emission in the high-$\chi$ case (Figure
\ref{fig:ISM}(b)). This is
because the ejecta is assumed to be cold for the neutron-free case,
while for the neutron-rich case, it is hot (see equation (\ref{Gamma_Ith})).
At the RRS front, the original thermal energy of the protons would
be also shared by the electrons, leading to stronger emission.

The neutron-fed signatures last only for a few 100 seconds. Rapid and
frequent monitoring (e.g. with $\sim$ 5 s exposure interval) of
early optical afterglow is needed to catch these signatures. Swift
UVOT, with an on-target time 60-100 s, has the capability to detect such
signals. 
 
\section{Early optical afterglow lightcurves: short
GRBs}\label{NumExap:Short} 

Bursts with durations shorter than 2 s may have different physical
origin. The leading model for short GRBs invokes mergers of two
compact objects (e.g., NS-NS merger or BH-NS merger, see e.g. Eichler
et al. 1989). Although there are other suggestions about the origin of
short GRBs, the merger model was found suitable to interpret many
short GRB properties (e.g. Ruffert et al. 1997). A pre-concept of such
a merger model is that the burst site is expected to have a large
offset from the host galaxy due to asymmetric kicks during the births
of the NSs, and that the number density of the external medium is low,
typically $0.01{\rm cm^{-3}}$. Such a suggestion seems to receive
support from the afterglow data of the recent short, soft GRB 040924
(Fan et al. 2005b). As already mentioned in the introduction, in the
compact object merger model (especially the NS-NS merger model), the
outflow is very likely to be neutron-rich ($\chi \geq 1$). Here we
calculate the early optical afterglow lightcurve of a neutron-rich
short GRB ($\chi=1$) and compare it with the neutron-free case. 

We take the following typical parameters for a short GRB. The burst
duration is taken $\sim 0.2{\rm s}$, so that the width of the
N-ejecta is $\Delta\sim 6\times 10^{9}{\rm cm}$. The total (include
both protons and neutrons) isotropic energy is $E_{\rm tot}=2.0\times
10^{51}{\rm ergs}$. The initial LF of the merged ion shell and the
neutron shell are $\Gamma_{\rm m}=200$, $\Gamma_{\rm n}=300$,
respectively\footnote{Notice that for a short GRB, as 
the I-ejecta crosses the slightly decayed neutron shell, the
interaction is so weak that there is no UV flash predicted for long
GRBs (Fan \& Wei 2004).}. The I-eject and the N-ejecta separate at a
radius $\approx 2\Gamma_{\rm m}^2\Gamma_{\rm n}^2\Delta/(\Gamma_{\rm
n}^2-\Gamma_{\rm m}^2)\approx 10^{15}{\rm cm}$, which is much smaller
than the typical decay radius $\simeq 8\times 10^{15}{\rm cm}$. As a
result, for short bursts most of the neutron decay products are
deposited onto the ISM (i.e. $f\sim 1$). This is the ``clean''
scenario discussed by B03. The resulting $\zeta(R)$ in the trail is
very large for $R\sim {\rm several}~R_{\beta}$, and a gap still forms
between the trail (T-ejecta) and the I-ejecta. The calculation is
therefore the same as the ISM-model for long GRBs (see \S{\ref{ISM}}
for detail). For $\Delta\sim 10^{10}{\rm cm}$ and $\Gamma_{\rm m}\sim
200$, the I-ejecta starts to spread at $R>10^{15}{\rm cm}$ (e.g.,
Piran 1999). This radius is much smaller than the deceleration
radius, which is $\simeq 7\times 10^{16}E_{\rm
tot,51}^{1/3}n_{-2}^{-1/3}\Gamma_{\rm m,2.3}^{-2/3}{\rm cm}$ for
$\chi=0$. In our calculation, we take the spreaded width $\sim 5\times 
10^{11}{\rm cm}$ when calculating the RS emission. Also here we
perform numerical calculations as compared with the analytical
treatment presented in Fan et al. (2005b). We found that the relative
LF between the unshocked I-ejecta and the shocked I-ejecta is smaller
than the one estimated in Fan et al. (2005b), so that the RS emission
is further suppressed.

\begin{figure}
\epsscale{1.00}
\plotone{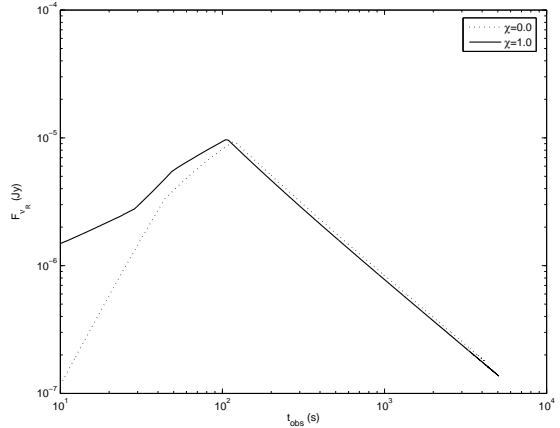}
\caption{Early R-Band lightcurves for a typical short
 GRB for $\chi=1$ (solid) and $\chi=0$ (dotted), respectively.
 The input parameters are the same as those listed in the caption of
 figure \ref{fig:WIND}, except that
 $n=0.01{\rm cm^{-3}}$ and $E_{\rm tot}=2\times 10^{51}{\rm ergs}$ are
 adopted.
\label{fig:Short}}
\end{figure}

The lightcurves are plotted in figure \ref{fig:Short} for $\chi=0$ and
$\chi=1$. Similar to figure \ref{fig:ISM}(b), the main difference
between the two cases is that for the latter the very early
lightcurve increases more slowly with time. This is due to the
contribution from the shocked trail. A collision also happens as the
I-ejecta catches up with the decelerated T-ejecta. The signature,
which is manifested as a steeper increasing lightcurve around 20 s, is
however not as prominent as the long GRB case. There are two
reasons. One is that the T-ejecta and the I-ejecta now have a
comparable mass, since essentially all the neutron decay products are 
stored in the T-ejecta (for comparison, for long bursts, a good
fraction is deposited onto the I-ejecta so that the I-ejecta is much
more energetic than the T-ejecta). Second, the RRS is very weak, and
the main contribution to the lightcurve is the FS component (for
comparison, for long bursts, the RRS component dominates the bump
emission). Nonetheless, we identify some novel features of a
neutron-rich short GRB. The signature occurs too early, however, and
the global afterglow level is very dim due to the small
$E_{\rm tot}$ and $n$. It is still a challenging task to diagnose the
neutron component in short GRBs.

\section{Discussion and summary}\label{Dis}

The composition of the outflow that powers GRBs is still far from
clear. The well-collected late afterglow data are unfortunately not
suitable to study the GRB fireball composition, since the afterglow
radiation comes from the shocked medium rather than the fireball
materials. Understanding the fireball composition and hence the nature
of the explosion requires detailed early afterglow data. In general a
GRB fireball is composed of two distinct components, a baryonic component
and a Poynting flux component. Current early afterglow data suggest
that GRB ejecta seem to be magnetized (e.g. Fan et al. 2002; Zhang et
al. 2003; Kumar \& Panaitescu 2003). Closer modeling suggests that
for GRB 990123 and GRB 021211 in which the reverse shock magnetic
field is stronger than that in the forward shock region, the $\sigma$
parameter, i.e. the Poynting flux to baryonic flux ratio, is moderate
(Zhang \& Kobayashi 2004; Fan et al. 2004b). In other two cases of
early afterglow (GRB 021004 and GRB 030418), detailed modeling is
needed to reveal whether Poynting flux is important. In any case, it
is likely that the baryonic component is at least non-negligible. As
discussed in the introduction, the neutron component is an inevitable
composition within the baryonic component, and the diagnosis of the
neutron component is a handle to reveal the significance of the
baryonic component in GRB fireballs. 
In the literature, the possible neutron signatures discussed include
the multi-GeV neutrino emission (e.g. Derishev et al. 1999a; Bachall \&
M\'{e}sz\'{a}ros 2000; 
M\'{e}sz\'{a}ros \& Rees 2000) and the UV flash accompanying the
$\gamma-$ray emission phase (Fan \& Wei 2004) \footnote{This prediction 
 may have already been confirmed by the recent detection of the 
 prompt optical and IR emission (which are variable and correlated with
 the $\gamma-$ray emission) during GRB 041219a (Blake et al. 2005; Vestrand et al. 2005). }. 
The detection of these signals are difficult due to the limitation of
the current detector capability. 

The most prominent neutron signature is likely imprinted in the early
afterglow. This suggestion has been proposed
(e.g. Derishev et al. 1999b; B03), but no detailed calculations have
been performed. In this paper, we present a first detailed calculation
of early optical afterglow lightcurves for a neutron-fed fireball. We
considered both long and short GRBs, and for long GRBs we consider
both a ISM medium and a wind medium. For each model, we calculate
the lightcurves for different $\chi$ (the neutron-to-proton number
density ratio) values, aiming to study how the neutron component
progressively change the lightcurve behavior as $\chi$ increases.
We confirmed the previous suggestions that
the presence of neutrons changes the fireball dynamics and the
lightcurves (Derishev et al. 1999b; B03). Our findings are summarized
as follows.

1. For short GRBs, the neutron decay products deposit onto the ISM
medium mostly, and the neutron decay trail clearly separates from
the ion ejecta (I-ejecta). This is the clean picture delineated in
B03. For long GRBs, the picture is more complicated. Only part of the
decay products are deposited onto the medium. The rest are deposited
onto the I-ejecta itself or onto the shocked region. The neutron
signatures in long GRBs therefore require more complicated treatments.

2. If the medium is a pre-stellar wind (for long GRBs), the neutron
trail moves slowly (mainly because the medium inertia is too
large). The trail and the I-ejecta do not separate from each other,
and a forward shock propagates into the trail directly. Three
components contribute to the final emission, i.e. the forward shock,
the reverse shock propagates into the I-ejecta, and the unshocked
trail emission. The latter is significant when $\chi$ is large, since
the internal energy of the unshocked trail is large when the medium
density is high. A typical neutron-rich wind-interaction lightcurve
is a characterized by a prominent early plateau lasting for $\sim 100$
s followed by a normal power-law decay (Fig.\ref{fig:WIND}). We also
show that in the wind case, the IC cooling effect due to the
overlapping of the initial prompt $\gamma-$ray with the shocks and the
trail (e.g. Beloborodov 2005; Fan et al. 2005a) suppresses the
very early R-band afterglow significantly, and the neutron-fed
signature is also dimmed  (see Fig.\ref{fig:WIND}(b) and
Fig.\ref{fig:WIND}(c) for a comparison).

3. If the medium is a constant density ISM (for long GRBs), part of
the neutron decay products fall onto the medium, and the trail moves
faster than the I-ejecta. A gap likely forms between the leading trail
and the I-ejecta. The former forms a distinct trail ejecta (T-ejecta)
which interacts with the out trail or ISM. The latter catches up later
and gives rise 
to a rebrigtening signature. Before collision, the radiation is
dominated by the forward shock emission. During the collision, both
the forward shock emission and the refreshed shocks (especially the
refreshed reverse shock) are important. The unshocked trail emission
is not important in this case. A typical neutron-rich ISM-interaction 
lightcurve is characterized by a slow initial rising lightcurve
followed by a prominent bump signature around tens to hundreds of seconds
(Fig.\ref{fig:ISM}). 

4. The picture for short GRBs is similar to the case of long GRBs with
ISM-interaction. The injection bump is less prominent and the
refreshed reverse shock is not important. A typical neutron-rich short
GRB lightcurve is characterized by a slow initial rising lightcurve
followed by a step-like injection signature (Fig.\ref{fig:Short}).

For all the cases, the predicted signatures can be detected by the
UVOT on board the {\em Swift} observatory. However, most of these
signatures (such as the plateau and the bump signature) are not
exclusively for neutron decay. More detailed modeling and case study are
needed to verify the existence of the neutron component.

A neutron-fed fireball involves very complicated processes and some
not well-studied physical problems (e.g. particle acceleration and
emission in the trail). In principle a complicated numerical model is
needed to delineate the problem. In this paper, we made some
approximations to simplify the problem and eventually come up with a
semi-analytical model. Such a treatment nonetheless catches the main
novel features of the model. Further studies are needed to build up
a more realistic model of neutron-fed fireballs.

So far there are four GRBs whose early afterglows are detected
(Akerlof et al. 1999; Fox et al. 2003a, 2003b; W. Li et al. 2003; Rykoff et
al. 2004, see Zhang \& M\'{e}sz\'{a}ros 2004 for a recent review), i.e. GRB
990123, GRB 021004, GRB 021211 and GRB 030418. Modeling GRB 990123 and
GRB 021211 suggests that these two bursts are born in an ISM medium
(e.g. Panaitescu \& Kumar 2002; Kumar \& Panaitescu 2003). The early
afterglow of these two bursts are dominated by fast-decaying optical
flash, usually interpreted as the reverse shock emission component. No
neutron-signature discussed in this paper has been discovered. There
are two possible reasons for this. First, the ejecta of these two
bursts are likely magnetized (Fan et al. 2002, 2004b; Zhang et
al. 2003;  Zhang \& Kobayashi 2004;
Kumar \& Panaitescu 2003; Panaitescu \& Kumar 2004; McMahon et al. 2004).
This would dilute the neutron signature discussed in this paper (in
which $\sigma=0$ has been assumed throughout). Second, modeling
suggests that the initial LF of the I-ejecta of these two bursts are
large, around 1000 (e.g. Wang et al. 2000; Soderberg \& Ramirez-Ruiz
2002; Wei 2003; Kumar \& Panaitescu 2003). Since the N-ejecta could
not be accelerated to a very high LF (Bahcall \& M\'esz\'aros 2000),
the N-ejecta likely lags behind the I-ejecta, and its influence on the
fireball dynamics is minimized.
The early afterglows of other two bursts GRB 021004 and GRB 030418 are
not easy to interpret within the standard reverse shock picture. One
suggestion is that both bursts are born in a wind environment (e.g. Li
\& Chevalier 2003; Rykoff et al. 2004). If this is the case, the 
available earliest detection at $\sim 300{\rm s}$ is too late to catch
the trail emission predicted in this paper. Alternatively, they are
potentially interpreted in an ISM model with the neutron signature
discussed in this paper. With the combination of the injection bump
and the forward shock bump (e.g. Kobayashi \& Zhang 2003a), one might
be able to achieve a broad early bump to interpret the early afterglow
behavior of GRB 021004 and GRB 030418. Finally, these two bursts may
be modeled by a high-$\sigma$ flow (e.g. Zhang \& Kobayashi 2004 for
more discussions). More detailed modeling is needed to verify these
suggestions. 

With the presence of magnetic fields, the acceleration and interaction
of the neutron component may have some novel features. In the
magnetization scenario, a two component jet is likely, with the wide
less collimated jet being due to the mildly relativistic neutrons
(Vlahakis et al. 2003; Peng et al. 2004). Peng et al. (2004) suggest
that the decay products in the wider neutron component would give rise
to an injection bump in the afterglow for an observer on-beam the
narrow core jet. In such a picture, for an observer off-beam of the
narrow jet but on-beam the wide jet, the observer would see an
orphan-afterglow-type event. Since the LF of the wide component is
about 15, the neutrons would decay at a typical distance of $\sim
4\times 10^{14}$ cm. The decay products shock into the ambient medium
and get decelerated at a further distance (the deceleration
distance). For somewhat optimistic parameters ($\epsilon_{\rm e}=0.3$,
$\epsilon_{\rm B}=0.1$, $p=2.3$, $n=1$ and $A_*=1/3$), the forward
shock typical synchrotron frequency at the typical neutron decay
radius is $\nu_{\rm m}\sim {\rm eV}$ for an ISM medium, or $\nu_{\rm
m}\sim 0.1~{\rm keV}$ for a wind-medium. In the ISM case, the optical
lightcurve increases rapidly with time and reaches a peak at the deceleration
time (typically hours after the burst trigger). In the wind case,
since the soft X-ray band is above both $\nu_m$ and $\nu_c$, the
lightcurve between the typical decay time and the deceleration time is
essentially flat ($\propto t^{(2-p)/2}$), giving rise to a flat soft
X-ray plateau lasting for hours. This would be an interesting
signature for the neutron-rich two-component jet model in the wind
environment. However, without a $\gamma$-ray or hard X-ray precursor,
detecting such a signature is challenging.

\acknowledgments We thank the referee for helpful suggestions to
improve the presentation of the paper. Y.Z.F. thanks T. Lu and Z. Li
for long term encouragement on the subject of neutron-fed GRBs, and
F. Peng for helpful discussion. We also thank P. M\'esz\'aros and
S. Kobayashi for reading the manuscript. This work is supported by
NASA NNG04GD51G and a NASA Swift GI (Cycle 1) program (for B.Z.),
the National Natural Science Foundation (grants 10073022,
10225314 and 10233010) of China, and the National 973 Project on
Fundamental Researches of China (NKBRSF G19990754) (for D.M.W.).

\appendix
\section{Appendix: Calculations of Synchrotron Radiation}\label{App-A}
Here we present the standard treatment of the synchrotron radiation of
electrons with a power-law energy distribution (e.g. Sari, Piran \&
Narayan 1998, and Piran 1999 for a
review). This is valid for all 
the cases invoking shocks, and it is also valid for one extreme model
of trail emission (as discussed in Appendix \ref{App-B}). As usual, we
introduce the equipartition parameters $\epsilon_{\rm e}\simeq 0.1$
and $\epsilon_{\rm B}\simeq 0.01$ as the energy fraction of the
electrons and magnetic fields in the local thermal energy in the
energy dissipation region (e.g. shock front), respectively.
The electrons are assumed to be distributed as a power law,
i.e. ${dN_{\rm e} / d\gamma_{\rm 
e}}\propto {\gamma}_{\rm e}^{\rm -p}~{\rm for}~\gamma_{\rm
e}>\gamma_{\rm e,m}$, where
\begin{equation}
\gamma_{\rm e,m}=\epsilon_{\rm e}(m_{\rm p}/m_{\rm e})[(p-2)/(p-1)]
[\gamma_{\rm r}(1+\gamma_{\rm th})-1]+1, 
\label{gamma_em}
\end{equation}
$m_{\rm p}$ and $m_{\rm e}$ are the rest mass of proton and electron respectively, $p\simeq 2.3$ is the
typical power law index of the electrons, $\gamma_{\rm r}$ is the
relative LF between the shock upstream and downstream, and
$\gamma_{\rm th}$ is the average random LF of protons in the
upstream. When deriving this expression, we have taken the downstream 
random LF as $\gamma_{\rm th,d}=\gamma_{\rm r} (1+\gamma_{\rm th,u})
-1$ (where $\gamma_{\rm th,u}=\gamma_{\rm th}$ in the above
expression), which is approximately valid when $\gamma_{\rm th,d} \gg
\gamma_{\rm th,u}$. The exact relativistic shock jump conditions 
for a hot upstream (Kumar \& Piran 2000; Zhang \& M\'esz\'aros 2002)
leads to  $\gamma_{\rm th,d}+1 \simeq (4/3) \gamma_{\rm r}
(1+\gamma_{\rm th,u})$ when $\gamma_{\rm th,d} \gg \gamma_{\rm th,u}$
is satisfied. So our treatment is valid to order of magnitude.

The co-moving magnetic field strength in the shock front can be
estimated by 
\begin{equation}
B'\approx\sqrt{32\pi \epsilon_{\rm B}\gamma_{\rm r}[\gamma_{\rm
r}(1+\gamma_{\rm th})-1]n_{\rm u}m_{\rm p}c^2}.
\end{equation}
where $n_{\rm u}$ is the co-moving proton number density in upstream. At a
particular time, there is a critical LF of the electrons,
\begin{equation}
\gamma_{\rm e,c}={6\pi
m_{\rm e}c\over (1+Y)\sigma_{\rm T }\Gamma {B'}^2t},
\label{gamma_ec}
\end{equation}
 above which the
electrons are cooled, $\Gamma$ is the bulk LF of the shocked region.
Here $\sigma_{\rm T}$ is the Thomson cross section, $Y$ is the
Compton parameter (e.g., Wei \& Lu 1998, 2000; Panaitescu \& Kumar 2000;
Sari \& Esin 2001; Zhang \& M\'{e}sz\'{a}ros 2001) which reads
$Y\simeq {-1+\sqrt{1+4x\epsilon_{\rm 
e}/\epsilon_{\rm B }}\over 2}$, where $x$ is the radiation coefficient
of the electrons, so that the total radiation efficiency is $\epsilon_{\rm
rad}\equiv x\epsilon_{\rm e}$, where $x=\min\{ 1,(1+k)(\gamma_{\rm
e,m}/\gamma_{\rm e,c})^{\rm (p-2)}/[k(3-p)]\}$, where $k=3,~1$ are for
the ISM and wind respectively. This last equation could be derived as
follows. 

Assuming that the outflow is in the slow cooling phase, one has $dN_{\rm
e}/d\gamma_{\rm e}=N_0(\gamma_{\rm e}/\gamma_{\rm e,m})^{\rm -p}$ for
$\gamma_{\rm e,m}<\gamma_{\rm e}<\gamma_{\rm e,c}$ and  $dN_{\rm
e}/d\gamma_{\rm e}=N_0\gamma_{\rm e,c}\gamma_{\rm
e,m}^{-1}(\gamma_{\rm e}/\gamma_{\rm e,m})^{\rm -(p+1)}$ for
$\gamma_{\rm e,c}<\gamma_{\rm e}<\gamma_{\rm e,M}$, where $\gamma_{\rm
e,M}\sim 10^8(B'/1\rm G)^{-1/2}\gg \gamma_{\rm e,c}$. Thus the total
emitting power (including the IC component) satisfies 
\begin{eqnarray}
P_{\rm tot}&=&(4/3)(1+Y)N_0\Gamma^2\sigma_{\rm
T}(B'^2/8\pi)c\gamma_{\rm e,m}^{\rm p}[\int^{\rm \gamma_{\rm
e,c}}_{\rm \gamma_{\rm e,m}} \gamma_{\rm e}^{\rm -(p-2)}d\gamma_{\rm
e}+\gamma_{\rm e,c}\int^{\rm \gamma_{\rm e,M}}_{\rm \gamma_{\rm e,c}}
\gamma_{\rm e}^{\rm -(p-1)}d\gamma_{\rm e}]\nonumber\\ 
&\approx&(B'^2/6\pi)(1+Y)N_0\Gamma^2\sigma_{\rm T}c\gamma_{\rm
e,m}^{\rm p}\gamma_{\rm e,c}^{\rm 3-p}/[(p-2)(3-p)], 
\label{P_tot}
\end{eqnarray}  
On the other hand, the total energy of the electrons reads
\begin{eqnarray}
W&=&N_0\Gamma\gamma_{\rm e,m}^{\rm p}m_{\rm e}c^2[\int^{\rm
\gamma_{\rm e,c}}_{\rm \gamma_{\rm e,m}} \gamma_{\rm e}^{\rm
-(p-1)}d\gamma_{\rm e}+\gamma_{\rm e,c}\int^{\rm \gamma_{\rm
e,M}}_{\rm \gamma_{\rm e,c}} \gamma_{\rm e}^{\rm -p}d\gamma_{\rm
e}]\nonumber\\ 
&\approx & N_0\Gamma\gamma_{\rm e,m}^2m_{\rm e}c^2/(p-2).
\label{W}
\end{eqnarray}  
The ratio of the fresh electrons ($d N_{\rm e}$) to the total
electrons ($N_{\rm e}$) satisfies 
\begin{equation}
d N_{\rm e}/N_{\rm e}\approx kdt/[(1+k)t],
\label{N_ratio}
\end{equation} 
where $k=3,~1$ are for the ISM and wind respectively. In deriving
equation (\ref{N_ratio}), relations $N_{\rm e}\propto R^{\rm k}$,
$dR\approx 4\Gamma^2cdt$ and $R\approx 4(1+k)\Gamma^2c t$ have been
used, and the outflow is assumed to be in the adiabatic
phase. So the total energy of the fresh electrons is $d W\approx W
dN_{\rm e}/N_{\rm e}$, the corresponding total input power
(including that of protons) is 
\begin{equation}
P_{\rm fresh}=dW/(\epsilon_{\rm e}dt)\approx kN_0\Gamma\gamma_{\rm
e,m}^2 m_{\rm e}c^2/[\epsilon_{\rm e}(1+k)(p-2)t], 
\end{equation}
Combining with equation (\ref{gamma_ec}), we have
\begin{equation}
x\equiv {P_{\rm tot}\over \epsilon_{\rm e} P_{\rm fresh}}= {(1+k)\over
k(3-p)}({\gamma_{\rm e,m}\over \gamma_{\rm e,c}})^{\rm p-2}.
\end{equation} 
This more precise result is slightly larger than that presented in
Sari \& Esin (2001). 

The typical synchrotron radiation frequency and the cooling frequency
are estimated by 
\begin{equation}
\nu_{\rm m}={\gamma_{\rm e,m}^2\Gamma eB'\over 2(1+z)\pi
m_{\rm e}c},
\end{equation}
and
\begin{equation}
\nu_{\rm c}={\gamma_{\rm e,c}^2\Gamma eB'\over 2(1+z)\pi
m_{\rm e}c},
\end{equation}
respectively, where $e$ is the electron charge, $z$ is the redshift of
the GRB. 

Another potentially important frequency involved (in the wind case) 
is the self-absorption frequency $\nu_{\rm a}$. Here we follow Rybicki
\& Lightman (1979, p.188-190) for a standard treatment: In the
co-moving frame of the emission region (all physical parameters
measured in co-moving frame are denoted with a prime),
the absorption coefficient $a'_{\nu'}$ scales as $a'_{\nu'}\propto
{\nu'}^{\rm -(p+4)/2}$ for $\nu'_{\rm b}\equiv \max(\rm
\nu'_m,\nu'_c)>{\nu'}>\nu'_{\rm p}\equiv
\min(\rm \nu'_m, \nu'_c)$, and as $a'_{\nu'}\propto
{\nu'}^{-5/3}$ for ${\nu'}<\nu'_{\rm p}$. 
For $\nu'_{\rm b}>{\nu'}>\nu'_{\rm p}$, one has
\begin{equation}
a'_{\nu'}=A{\nu'}^{\rm -(\bar p+4)/2},
\label{nu_m<nu_c}
\end{equation}
while for ${\nu'}<\nu'_{\rm p}$, one has 
\begin{equation}
a'_{\nu'}=A{\nu'_{\rm p}}^{\rm -(\bar p/2+1/3)}{\nu'}^{-5/3}.
\label{nu_c<nu_m}
\end{equation}
Here $A=\frac{\sqrt{3}e^3}{8\pi m_{\rm e}}\left(\frac{3e}{2\pi m_{\rm
e}^3c^5}
\right)^{\bar p/2}(m_{\rm e}c^2)^{\rm \bar p-1}KB'^{\rm (\bar p+2)/2}
\Gamma\left(\frac{3\bar p+2}{12}\right)\Gamma\left(\frac{3\bar
p+22}{12}\right)$, 
where $K\approx 2(\bar p-1)n_{\gamma_{\rm e}}{ \gamma_{\rm p}^{\rm
\bar p-1}}/3$, $n_{\gamma_{\rm e}}$ is the downstream proper proton number
density.  ${\gamma_{\rm p}\equiv\min(\gamma_{\rm e,m},\gamma_{\rm
e,c})}$, $\Gamma({\rm x})$ is the Gamma function. The above general
treatment is valid when we take $\bar p=p$ for the slow cooling and
$\bar p=2$ for the fast cooling. The self-absorption optical depth can
be calculated by 
\begin{equation}
\tau(\nu')=\int a'_{\rm \nu'}dR'.
\end{equation}
When considering shocks, we can approximate $\int a'_{\nu'}dR'\approx
a'_{\nu'}\Delta  R'$, where $\Delta R'$ is the width of the shocked
region, so that $n_{\gamma_{\rm e}}\Delta R'\simeq N_{\rm e}/4\pi
R^2$, where $N_{\rm e}$ is the total number of emitting electrons. The
co-moving self-absorption frequency $\nu'_{\rm a}$ can be derived from
$\tau(\nu'_{\rm a})=1$, and the observed absorption frequency is
$\nu_{\rm a}=\Gamma \nu'_{\rm a}/(1+z)$.

The synchrotron flux as a function of observer frequency can be
approximated as a four-segment broken power law. For $\nu_{\rm
a}<\min\{\nu_{\rm c},\nu_{\rm m}\}$, see Eqs. (4-5) of
Zhang \& M\'{e}sz\'{a}ros (2004).  For $\nu_{\rm
a}>\min\{\nu_{\rm c},~\nu_{\rm m}\}$, one has 
\begin{equation}
 F_{\rm \nu}\approx{F_{\rm max}} \left \{ \begin{array}{llll}
 {({\nu_{\rm c}/\nu_{\rm a}})^{3}(\nu/\nu_{\rm c})^{2}}, \,\,\,\, &
 (\nu<\nu_{\rm c}),\\ {({\nu_{\rm a}/\nu_{\rm c}})^{-1/2}(\nu/\nu_{\rm
 a})^{5/2}}, \,\,\,\, & (\nu_{\rm c}<\nu<\nu_{\rm a}),\\
 ({\nu/\nu_{\rm c}})^{-1/2},\, \,\,\,\, & (\nu_{\rm a}<\nu<\nu_{\rm
 m}), \\ ({\nu_{\rm m}/\nu_{\rm c}})^{-1/2}({\nu/\nu_{\rm m}})^{\rm
 -p/2},\,\,\,\,\, & (\nu>\nu_{\rm m}).  \end{array} \right.
 \label{nu_a1}
\end{equation}
for the fast cooling case, and
\begin{equation}
 F_{\rm \nu}\approx{F_{\rm max}} \left \{ \begin{array}{llll}
 ({\nu_{\rm m}\over \nu_{\rm a}})^{\rm {(p+4)\over 2}}({\nu\over
 \nu_{\rm m}})^{2}, \,\,\,\, & (\nu<\nu_{\rm m}),\\ ({\nu_{\rm a}\over
 \nu_{\rm m}})^{-\rm {(p-1)\over 2}}({\nu\over \nu_{\rm a}})^{5/2},
 \,\,\,\, & (\nu_{\rm m}<\nu<\nu_{\rm a}),\\ ({\nu\over \nu_{\rm
 m}})^{\rm -{(p-1)\over 2}},\, \,\,\,\, & (\nu_{\rm a}<\nu<\nu_{\rm
 c}), \\ ({\nu_{\rm c}\over \nu_{\rm m}})^{\rm -{(p-1)\over
 2}}({\nu\over \nu_{\rm c}})^{\rm -{p\over 2}},\,
\,\,\,\, & (\nu>\nu_{\rm c}),
   \end{array} \right.  \label{nu_a2}
\end{equation}
for the slow cooling case. 
Here $F_{\rm max}\approx {3\sqrt{3}\Phi_{\rm p}(1+z)N_{\rm e} m_{\rm
e}c^2\sigma_{\rm T}\over 32\pi^2 eD_{\rm L}^2}\Gamma B$, $\Phi_{\rm
p}$ is a function of $p$ (e.g. for $p\simeq 2.3$, $\Phi_{\rm p}\simeq
0.6$) (Wijers \& Galama 1999), and $D_{\rm L}$ being the luminosity
distance. 

\section{Appendix: Trail emission}\label{App-B}
One novel feature of the neutron-fed GRB afterglow, in particular in the
wind-medium case, is the emission from the neutron decay trail itself
(without being shocked). This is also one of the main challenges faced in
preparing this work. For example, it is unclear (1) what is the
mechanism of trail emission, (2) how much thermal energy is
distributed to electrons and magnetic fields, and (3) how the emitting
electrons are distributed.

The first two uncertainties also apply to the shock acceleration
case. Our treatment here closely follows the standard shock model. We
first assume that the trail emission mechanism is also synchrotron
radiation, and assign two equipartition parameters, i.e.
$\epsilon_{\rm e}\sim 0.1$ and $\epsilon_{\rm B}\sim 0.01$ 
for the electrons and the magnetic fields, respectively. For the 
third uncertainty, while there is a standard paradigm (i.e. Fermi
acceleration) in the shock case so that a power-law distribution of
electron energy is justified, it is unclear whether it is the case
in the trail. Rather than speculating the possible electron
distributions, in this paper we consider two extreme situations, which
may bracket the more realistic electron distribution case. In the
first model, we still consider a pure power-law model that is
completely analogous to the shock case, i.e. for an average electron
LF $\bar{\gamma}_{\rm e}$, we get the minimum electron LF $\gamma_{\rm
e,m}=[(p-2)/(p-1)] \bar{\gamma}_{\rm e}$, and $d N_e/d \gamma_{\rm e}
\propto (\gamma_{\rm e}/\gamma_{\rm e,m})^{-p}$ for $\gamma_{\rm e} >
\gamma_{\rm e,m}$. In the second model, the electrons are assumed to
be mono-energetic, i.e. all electrons have a LF of
$\bar{\gamma}_{\rm e}$.
A more realistic version of the mono-energetic model is the
relativistic Maxwellian distribution model, i.e. $dN_{\rm
e}/d\gamma_{\rm e}\propto \gamma_{\rm e}^2\exp(-\gamma_{\rm e}m_{\rm
e}c^2/kT_{\rm e})$, where $T_{\rm e}$ is the temperature of the
plasma. For an observer frequency far above several $\times
\bar{\nu}_{\rm e}$, where $\bar{\nu}_{\rm e}=\gamma(R) \bar{\gamma}_{\rm
e}^2eB'/2\pi (1+z)m_{\rm e}c$, the observed flux for both the
mono-energetic model and the Maxwellian model is much dimmer than the
one in the power law model, where $\gamma(R)$ is the bulk LF of the trail. 
Below $\bar{\nu}_{\rm e}$, on the other hand, the synchrotron emission
properties are quite similar for various distribution models. 
Fortunately, for the wind-interaction case when the trail emission
becomes important, 
$\bar{\nu}_{\rm e}$ is indeed above the R-band frequency $\nu_{\rm
R}=4.6\times 10^{14}{\rm Hz}$, so that the uncertainty to calculate
the trail emission lightcurve is small (see Fig.\ref{fig:Trail}). 

The magnetic field generated in the neutron-front can be estimated by
\begin{equation}
B_{\rm tr}'\approx\sqrt{8\pi \epsilon_{\rm B}\gamma_{\rm th}n_{\rm tr}m_{\rm
p}c^2},
\label{B_tr}
\end{equation}
where $n_{\rm tr}$ has been shown in equation (\ref{B0303}). 
Please notice the different expression for $B'$ and $B'_{\rm tr}$.
We also assume $\epsilon_{\rm e}\sim 0.1$ in the trail, which may be
conservative. Even with this estimate, the trail emission is found to
be strong enough to be detectable in the wind case (see
Fig.\ref{fig:WIND} for detail).

A self-consistent calculation for the trail emission is quite
complicated. In this work we make the following approximations. (i)
After the medium is ``ignited'' by the neutron front at 
\begin{equation}
t_{\rm ign}=R/2\Gamma_{\rm n}^2c,
\label{t_ign}
\end{equation}
the trail located at $R$ (but moving with $\gamma(R)$) continually
contributes to the observed flux until it is ``terminated'' by the
I-ejecta shock front. (ii) The
trail is divided into many sub-layers, each moving with $\gamma(R)$
without interaction. The observed trail emission is a sum of the
independent radiation from these sub-layers. During the ``lifetime'' of
each sub-layer, the thermal LF of a particular electron cools as
$\gamma_{\rm e}(t)=\gamma_{\rm 
e,0}/[1+(1+Y)(\sigma_{\rm T}{B'_{\rm tr}}^2/6\pi m_{\rm
e}c)\gamma_{\rm e,0}\gamma(R)(t-t_{\rm ign})]$, where $\gamma_{\rm
e,0}$ is the initial LF of the electron. 
For the power law model, the electrons are assumed to be distributed
as a broken power law (considering cooling), but equations
(\ref{gamma_em}) and (\ref{gamma_ec}) are now replaced by 
\begin{equation}
\gamma_{\rm
e,m}(t)=\gamma_{\rm e,m}/[1+(1+Y)(\sigma_{\rm T}{B'_{\rm tr}}^2/6\pi 
m_{\rm e}c)\gamma_{\rm e,m}\gamma(R)(t-t_{\rm ign})]
\label{gamma_emt}
\end{equation} 
and
\begin{equation}
\gamma_{\rm e,c}(t)=6\pi m_{\rm e} c/(1+Y)[\sigma_{\rm T}{B'_{\rm
tr}}^2\gamma(R)(t-t_{\rm ign})],
\label{gamma_ect}
\end{equation}
respectively.

In the ISM case, for $n_{_{\rm ISM}}=1~{\rm cm}^{-3}$ (the favored
value for the current multi-wavelength afterglow modeling), the trail
emission never becomes dominant in the early R-band afterglows. This
is contrary to the wind case, in which $\zeta(R)$ is only moderate
while $\gamma_{\rm th}\gg 1$, even at $R\ll R_{\beta}$. The
synchrotron radiation from the trail is therefore strong, which
contributes significantly to the early R-band afterglow lightcurves
(see Fig.\ref{fig:Trail}).

\end{document}